\documentclass[iop,apj]{emulateapj}
\usepackage{color}
\usepackage{graphicx}
\usepackage{rotating}
\usepackage{amsmath}
\usepackage{longtable}

\newcommand{\twelveoh}{\ensuremath{12+\log[\textrm{O}/\textrm{H}]}}
\newcommand{\htwo}{\ensuremath{\textrm{H}_2}}

\begin{document}

\title{LIGHTING THE DARK MOLECULAR GAS: \htwo\ AS A DIRECT TRACER}
\author{Aditya Togi\altaffilmark{1} and J. D. T. Smith\altaffilmark{1}}
\altaffiltext{1}{Ritter Astrophysical Research Center, University of Toledo, 2825 West Bancroft Street, M. S. 113, Toledo, OH 43606\\
Email: aditya.togi@utoledo.edu}

\begin{abstract}
Robust knowledge of molecular gas mass is critical for understanding star formation in galaxies. The \htwo\ molecule does not emit efficiently in the cold interstellar medium, hence the molecular gas content of galaxies is typically inferred using indirect tracers.  At low metallicity and in other extreme environments, these tracers can be subject to substantial biases. We present a new method of estimating total molecular gas mass in galaxies directly from pure mid-infrared rotational \htwo\ emission.  By assuming a power-law distribution of \htwo\ rotational temperatures, we can accurately model \htwo\ excitation and reliably obtain warm ($T\!\gtrsim\!100$\,K) \htwo\ gas masses by varying only the power law's slope.   With sensitivities typical of Spitzer/IRS, we are able to directly probe the \htwo\ content via rotational emission down to $\sim80$\,K, accounting for $\sim15\%$ of the total molecular gas mass in a galaxy.  By extrapolating the fitted power law temperature distributions to a calibrated \emph{single} lower cutoff temperature, the model also recovers the total molecular content within a factor of $\sim$2.2 in a diverse sample of galaxies, and a subset of broken power law models performs similarly well.  In ULIRGs, the fraction of warm \htwo\ gas rises with dust temperature, with some dependency on $\alpha_\mathrm{CO}$. In a sample of five low metallicity galaxies ranging down to $\twelveoh=7.8$, the model yields molecular masses up to $\sim 100\times$ larger than implied by CO, in good agreement with other methods based on dust mass and star formation depletion timescale.  This technique offers real promise for assessing molecular content in the early universe where CO and dust-based methods may fail.
\end{abstract}


\section{Introduction \label{sec:intro}}
By a factor of more than $10^4$, \htwo\ is the most abundant molecule in the universe, found in diverse environments ranging from planet atmospheres to quasars hosts \citep{Hanel79, Hanel81, Walter03, Genzel14}.  It was the first neutral molecule to form in the Universe and hence dominated the cooling of pristine gas at early times \citep{Lepp02}.  Stars form principally from molecular clouds, and most physical prescriptions for stellar formation rate (SFR) are therefore directly linked to the surface density of \htwo\ gas \citep{Kennicutt98, Bigiel08}.  To understand the star formation processes and how it varies and has evolved over the star-forming history of the Universe, it is necessary to accurately measure the mass and distribution of \htwo.

Despite its high abundance, \htwo\ can be difficult to study directly.  It possess no permanent dipole moment, which makes it a weak rotational emitter.  In addition, the upper energy level for the lowest permitted rotational (quadrupole) transition is $E/k=510$ K above ground \citep{Dabrowski84, Dishoeck86}.  Hence, the \htwo\ gas that comprises the bulk of the molecular interstellar medium (ISM) is commonly believed to be too cold to be visible. And yet, despite these observational disadvantages, the advent of the Infrared Space Observatory (ISO) and in particular Spitzer's Infrared Spectrograph \citep[IRS,][]{Houck04} revealed pure rotational \htwo\ emission from a rich variety of extragalactic sources, including normal star forming
galaxies \citep{Roussel07}, ultraluminous and luminous infrared galaxies \citep[U/LIRGs][]{Lutz03, Pereira10, Veilleux09, Stierwalt14}, galaxy mergers \citep{Appleton06}, radio-loud AGN \citep{Ocana10}, UV-selected galaxies \citep{ODowd09}, quasar hosts \citep{Evans01}, cooling-flow cluster systems \citep{Egami06}, sources with extreme shock-dominated energetics \citep{Ogle10}, and even star-forming sources at redshifts $\gtrsim 2$ \citep{Ogle12}.

In the absence of direct measurements of \htwo, the lower rotational line transitions of CO (the next most abundant molecule) are generally used as a molecular gas tracer.  To convert the measured integrated intensities of $^{12}$CO to \htwo\ column density a conversion factor,  $\rm{\alpha_{CO}}$, is needed, typically calibrated against virial mass estimates in presumed self-gravitating clouds \citep{Solomon87, Scoville87, Strong96, Abdo10}.  However recent evidence indicates that the value of $\rm{\alpha_{CO}}$ varies substantially, both within galaxies and at different epochs \citep{Genzel12, Sandstrom13}.

In molecular clouds in our Galaxy, the observed $\gamma$-ray flux arising from the cosmic ray interactions with \htwo\ can be used to recover \htwo\ gas mass accurately \citep{Bhat85, Bloemen84, Bloemen86}, but this method requires information on the cosmic ray distribution not available in other galaxies.  Another method to assess \htwo\ content is to use dust as a tracer, with a presumed or recovered dust-to-gas (DGR) ratio \citep[e.g.][]{Sandstrom13, Remy14}. Among other biases, both the CO and dust-based methods lose reliability at low metallicity.  Due to declining dust opacity and less effective self-shielding than \htwo, the CO abundance drops rapidly with decreasing metallicity \citep{Wolfire10, Bolatto13}.  Variations in the radiation field can also lead to selective CO destruction \citep{Hudson71, Bolatto13}.  At the lowest metallicities, DGR itself begins to scale non-linearly with the metal content of the gas \citep{Herrera12, Fisher14, Remy14}.  Another method of inferring molecular gas content is to couple measured star formation rates with an assumed constant \htwo\ depletion timescale ($\tau_{dep}$), equivalent to a constant star formation efficiency (SFE) in galaxies \citep{Schruba12}.  None of these methods provide \emph{direct} tracers of the \htwo\ reservoirs, and the relevant indirect tracers all rely on physical assumptions (e.g. constant or measurable values of $\rm{\alpha_{CO}}$, DGR, or $\tau_{dep}$) which are unlikely to be valid in all environments.

Here we challenge the long-stated assumption that \htwo\ rotational emission is a poor tracer of the total molecular content in galaxies. Our understanding of the structure of the molecular material in galaxies is undergoing significant revision. While CO traces the coldest component of the molecular ISM, half or more (and sometimes much more) of the molecular gas in galaxies is in a warmer, CO-dark state, where \htwo\ persists \citep{Field66, Tielens05, Draine11, Wolfire10, Pineda13, Velusamy14}.  What this means is that the common wisdom that all molecular gas is found at temperatures $T=10-20$\,K is incorrect.  With molecular material existing in quantity at excitation temperatures 50 - 100\,K, it becomes possible to use the rotational emission spectrum of the Universe's dominant molecule to directly assess a substantial portion of the total molecular content in galaxies.

We introduce here a simple continuous temperature model which enables direct use of the \htwo\ rotational emission from galaxies to recover their total molecular gas content.  While several assumptions are required to make such a model possible, the resulting biases are expected to be completely distinct from those of the indirect gas tracers. The paper is laid out as follows. We describe the archival sample in \S\,2. A rotational temperature distribution model is presented in \S\,3, and in \S\,4 we describe the method and its calibration procedure. Section 5 presents results, discussions, applications and future prospects of our model. We summarize our conclusions in \S\,6.

\section{Sample}
We have selected a large sample of galaxies spanning a wide range of different physical properties with reliable \htwo\ rotational emission detections. Our sample includes 14 Low Ionization Nuclear Emission Regions (LINERs), 18 galaxies with nuclei powered by star formation, 6 Seyfert galaxies, 5 dwarf galaxies, 11 radio galaxies, 19 Ultra Luminous Infrared galaxies (ULIRGs), 9 Luminous Infrared galaxies (LIRGs), and 1 Quasi Stellar Object (QSO) host galaxy. The galaxies were chosen to have at least three well-detected \htwo\ pure-rotational lines with Spitzer/IRS \citep{Houck04}, including either S(0) or S(1), occurring at 28.22 and 17.04 $\micron$, respectively.   We also required an available measurements of CO line emission (either J = 1--0 or 2--1) for an alternative estimate of the total molecular gas mass.  See Table \ref{table:sample} for a complete sample list with relevant physical parameters and the references from which each galaxy was drawn.

The flux ratios $S_{70}/S_{160}$ are calculated using 70 and 160$\micron$ intensities obtained from the \emph{Herschel-PACS} instrument \citep{Poglitsch10}. We convolved PACS 70 $\micron$ maps to the lower resolution of PACS 160 $\micron$ to calculate flux ratios for mapped regions, for which we have the \htwo\ line flux. For ULIRGs  and other unresolved galaxies, $S_{70}/S_{160}$ represents the global flux ratio. The PACS70/PACS160 ratios are listed in Table \ref{table:sample} for each galaxy.

\subsection{MIR \htwo\ rotational line fluxes}
All MIR \htwo\ rotational line fluxes were obtained from \emph{Spitzer-IRS} observations at low (R $\sim$ 60 -- 120) and high (R $\sim$ 600) resolution between 5 -- 38 $\mu$m.  The Spitzer Infrared Nearby Galaxies Survey \citep[SINGS,][]{Kennicutt03} is a diverse sample of nearby galaxies spanning a wide range of properties. The SINGS sample consists of LINERs, Seyferts, dwarfs, and galaxies dominated by star formation in their nuclei.  We adopted the \htwo\ rotational line fluxes for four lowest rotational lines (S(0) to S(3)) in SINGS galaxies from \citet{Roussel07}. The targets were observed in spectral mapping mode and the details of the observing strategy is described in \citet{Kennicutt03} and \citet{Smith04}.

\htwo\ rotational line fluxes for 10 3C radio galaxies are from \citet{Ogle10}. This sample includes from S(0) to S(7).  For the radio galaxy 3C\thinspace236, the fluxes of \htwo\ rotational lines are obtained from \citet{Guillard12}.

The \htwo\ rotational line fluxes for QSO PG\thinspace1440+356 were taken from the Spitzer Quasar and ULIRG Evolution Study (QUEST) sample of \citet{Veilleux09}, and those for the 19 ULIRGs were obtained from \citet{Higdon06}. The fluxes of \htwo\ rotational lines for the NGC\thinspace6240 and other LIRGs in the sample are from \citet{Armus06} and \citet{Pereira10}, respectively.  \htwo\ rotational lines are also observed in the molecular outflow region of NGC\thinspace1266 \citep{Alatalo11}.  Table \ref{table:h2flux} lists the flux of MIR \htwo\ rotational lines for all our selected galaxies.

\subsection{Cold molecular gas mass from CO line intensities}
To test and calibrate a model which measures total molecular mass from the rotational lines of \htwo, an estimate of the ``true'' cold \htwo\ gas mass in a subset of galaxies is required. For this purpose, we utilize the cold \htwo\ gas masses for the SINGS sample compiled by \citet{Roussel07}, which are derived from aperture-matched CO velocity integrated intensities.  The observed line intensities of the 2.6 mm $^{12}$CO(1--0) line within the measured CO beam size was chosen to match the aperture of the \emph{Spitzer-IRS} spectroscopic observations. \citet{Roussel07} assumed a CO intensity to molecular gas conversion factor ($\rm{\alpha_{CO}}$) of 5.0 M$_\odot$(K km s$^{-1}$ pc$^{2}$)$^{-1}$ (equivalent to $\rm{X_{CO}} = 2.3\times 10^{20}$ cm$^{-2}$(K km s$^{-1}$)$^{-1}$). They applied aperture corrections by projecting the IRS beam and CO beam on the 8 $\micron$ \emph{Spitzer-IRAC} map. The underlying assumption is that the spatial distribution of the aromatic band in emission and CO(1--0) line emission are similar at large spatial scales.

For the radio galaxies, molecular gas masses were compiled by \citet{Ogle10}, and the molecular gas mass was estimated from the CO flux density corrected and scaled to the standard Galactic CO conversion factor. For the radio galaxy 3C\thinspace236 Oca{\~n}a, Flaquer et al. (2010) estimated an upper limit to the cold \htwo\ gas mass. The cold \htwo\ gas masses for the LIRGs in our sample were obtained from \citet{Sanders91}. They used $\rm{\alpha_{CO}}$ = 6.5 M$_\odot$(K km s$^{-1}$ pc$^{2}$)$^{-1}$ (equivalent to $\rm{X_{CO}} = 3.0\times 10^{20}$ cm$^{-2}$(K km s$^{-1}$)$^{-1}$). We derived aperture corrections to the CO intensities by projecting IRS and CO beam on the 8 $\micron$ IRAC map, since the CO beam size (55$\arcsec$ aperture) is much larger than the IRS mapped region ($13.4\arcsec\times13.4\arcsec$). Required aperture corrections are typically less than a factor of 2.  For the LIRGs NGC\thinspace7591, NGC\thinspace7130, and NGC\thinspace3256, the value of molecular gas mass was obtained from \citet{Lavezzi98, Curran00, Sakamoto06}, respectively. The global molecular gas masses for ULIRGs are from \citet{Rigopoulou96, Sanders91, Mirabel89, Evans02}.  For the quasar QSO PG\thinspace1440+356 \citet{Evans01} derived the molecular gas mass.

Different studies have utilized substantially different $\rm{\alpha_{CO}}$ values to calculate molecular gas masses.  In the Milky Way disk, a modern value of $\rm{\alpha_{CO,Gal}} = 4.35\pm1.3$  M$_\odot$(K km s$^{-1}$ pc$^{2}$)$^{-1}$ (equivalent to $\rm{X_{CO}} = 2.0\times 10^{20}$ cm$^{-2}$(K km s$^{-1}$)$^{-1}$) has been found to be consistent with a wide variety of constraints \citep{Bolatto13}.  In this work, we scale all molecular gas mass estimates to correspond to $\rm{\alpha_{CO}} = \rm{\alpha_{CO,Gal}}$. Since we require only the \htwo\ gas masses in our analysis, the resulting molecular mass values were further reduced by a factor of 1.36 to remove the mass contribution of Helium and other heavy elements.  Recent dust-based studies of a subset of the SINGS sample have revealed variations in $\rm{\alpha_{CO}}$ both between and within these galaxies \citet{Sandstrom13}.

Table \ref{table:h2flux} lists the CO-based \htwo\ gas mass estimated for each galaxy in the sample, aperture matched to the region of Spitzer/IRS coverage, along with \htwo\ masses calculated using the conversion factor for central regions derived by \citet{Sandstrom13} in parentheses, where available.

\section{Model}
\label{sec:model}
The primary methods of measuring \htwo\ mass all rely on a set of assumptions --- that the $\rm{\alpha_{CO}}$ factor is known and constant, that the dust-to-gas ratio is fixed or tied directly to metallicity, or that the star formation depletion time is unchanged from environment to environment.  All of these methods also rely on indirect tracers: the CO molecule, which is 10,000$\times$ less abundant than \htwo, dust grains, with their complex formation and destruction histories and varying illumination conditions, or newly formed stars, which are presumed to be associated with molecular material with a consistent conversion efficiency.  All of these methods suffer biases based on the physical assumptions made.

We propose a \textit{direct} means of assessing total molecular gas mass using the MIR quadrapole rotational line emission of \htwo.  Interstellar or direct ultraviolet radiation fields, shocks, and other mechanical heating processes are all potential sources of \htwo\ excitation.  The method we propose requires only the assumption that \htwo\ rotational temperatures are, when averaged over the diversity of emitting environments and processes in galaxies, widely distributed, and smoothly varying.  This is directly analogous to modeling the radiation field intensity which heats dust grain populations in galaxies with a smooth distribution \citep[e.g.][]{Draine07, Galliano03}.  We model the \htwo\ temperature distribution as a smooth, truncated \textbf{power law} with fixed cutoff temperatures.    Indeed, the \htwo\ molecule is radiatively cooled by a cooling function which can be well approximated as a power law with slope 3.8 \citep{Hollenbach79, Draine93}. A similar analysis by \citet{Burton87} derived a cooling function with a power law index, n = 4.66, for \htwo\ molecules in the temperature range 10--2000 K.

The temperature equivalents of \htwo\ rotational levels are high, starting at T = 510 K (see Table \ref{table:lineppt}). The Boltzmann distribution of energy levels, however, leads to substantial excitation even at more modest peak temperatures.  The lowest upper energy levels at J = 2--4 are well populated at excitation temperatures of 50 to 150 K, whereas the higher-J levels are excited by temperatures from a few hundred to several hundred K.  Typically, a small number of discrete temperature components are fitted to low-J rotational line fluxes, recovering warm \htwo\ temperatures in this range \citep{Spitzer73, Spitzercoch73, Spitzer74, Spitzerzweibel74, Savage77, Valentijn99, Rachford02, Browning03, Snow06, Ingalls11}.

Other studies have also adopted continuous temperature distributions for \htwo. \citet{Zakamska10} in her study of \htwo\ emission in ULIRGs used a power law model $dM \propto T^{-n}dT$ to calculate the expected \htwo\ rotational line ratios, where $dM$ is the mass of \htwo\ gas with excitation temperatures between $T$ and $T+dT$. In her sample,  power law indices $2.5<n<5.0$ are required to reproduce the observed range of \htwo\ line ratios. Similarly \citet{Pereira14} adopted a power-law distribution in \htwo\ to model \htwo\ rotational emission in six local infrared bright Seyfert galaxies, with power law indexes ranging from 4--5. In the shocked environment of supernova remnant IC 443, \citet{Neufeld08} found  a power law temperature distribution for \htwo\ column density with power law index varying over the range 3--6 with an average value of 4.5. 

The present practice for measuring warm ($T\gtrsim 150$K) \htwo\ gas mass is to use two or three distinct temperature components to model the \htwo\ rotational line fluxes.  Yet even on the scales of individual molecular clouds, \htwo\ is present at a wide range of temperatures, calling into question whether discrete temperature components are physical.  When used purely for assessing warm gas mass (T $\gtrsim$ 100K), the power law method provides a more robust, unique and reproducible measure than discrete temperature fits, which are sensitive to the arbitrary choice of starting line pair.  A continuous distribution also makes possible extrapolation to a suitable lower temperature to recover the total cold molecular gas mass (see \S\,\ref{sec:method--calibration}).

We fit the flux of \htwo\ rotational lines using a continuous power law temperature distribution by assuming
\begin{equation}
dN = m T^{-n} dT,
\label{eqn:powerlaw}
\end{equation}
\noindent where $dN$ is the column density of \htwo\ gas between excitation temperature $T$ and $T+dT$, $n$ is the power law index, and $m$ is a constant. Integrating this distribution to recover the total column density, the scaling co-efficient $m$ is found to be

\begin{equation}
m = \frac{N_{tot}(n-1)}{T_{\ell}^{1-n}-T_{u}^{1-n}},
\label{eqn:ntot}
\end{equation}
\noindent where $T_{\ell}$ and $T_{u}$ are the lower and upper temperatures of the distribution, respectively, and N$_{obs}$ is the column density of $H_{2}$ in the observed line of sight and temperature is presumed equal to the rotational temperature. For a continuous distribution of molecules with respect to temperature, the column density of molecules at upper energy level $j$ responsible for the transition line $S(j)$ is

\begin{equation}
N_{j} = \int_{T_{\ell}}^{T_{u}} \frac{g_{j}}{Z(T)} \times e^{\frac{-E_{j}}{kT}} m T^{-n} dT,
\label{eqn:nuj}
\end{equation}
\noindent where $g_{j}$ is the degeneracy value for the corresponding energy level $E_{j}$. The degeneracy value for even and odd values of $j$ corresponding to para and ortho \htwo\ is

\begin{equation}
g_{j} = 2j+1,
\end{equation}
and
\begin{equation}
g_{j} = 3(2j+1), 
\end{equation}
respectively. The factor of 3 for odd values of $j$ is for ortho-hydrogen, which have parallel proton and electron spins, forming a triplet state. $Z(T)$ is the partition function at temperature $T$. The model assumes local thermodynamic equilibrium (LTE) between ortho and para $H_{2}$ and the abundance ratio of ortho to para $H_{2}$ at any temperature, $T$, in LTE is given by \citet{Burton92}.  The partition functions for para and ortho \htwo\ are
\begin{equation}
Z(T) = Z_{p}(T) = \sum_{j=even}^\infty (2j+1) e^{\frac{-E_{j}}{kT}}
\end{equation}
and
\begin{equation}
Z(T) = Z_{o}(T) = \sum_{j=odd}^\infty 3(2j+1) e^{\frac{-E_{j}}{kT}} ,
\end{equation}
\noindent respectively. 

Table \ref{table:lineppt} list the wavelength of each \htwo\ rotational line transition, the upper energy level of the corresponding transition in temperature units, and their radiative rate coefficients $A$ \citep{Huber79, Black76}.\\

Assuming \htwo\ emission to be optically thin, the flux observed in a given transition $j$ is 
\begin{equation}
F_{j} = \frac{h\nu A N_{j+2} \Omega}{4\pi},
\label{eqn:fl}
\end{equation}
\noindent where $\Omega$ is the solid angle of the observation. Substituting the value of $N_{j}$  from equation \ref{eqn:nuj} into Eq.~\ref{eqn:fl} and then using equation \ref{eqn:ntot}, we obtain the total column density
\begin{equation}
N_{tot} = \frac{4\pi F_{j} \lambda (T_{\ell}^{1-n}-T_{u}^{1-n})}{A   hc \Omega (n-1) \int_{T_{\ell}}^{T_{u}}\frac{g_{j+2}}{Z(T)} e^{\frac{-E_{j+2}}{kT}} T^{-n} dT }
\label{eqn:ntot2} 
\end{equation}
\noindent From Eq.~\ref{eqn:fl}, we conclude that the column density is proportional to the flux $F$ and the corresponding transition wavelength $\lambda$, and inversely proportional to the spontaneous emission probability $A$,

\begin{equation}
N_{j+2} \propto \frac{F_{j} \lambda}{A}
\end{equation}
\noindent To perform the fit, we develop an excitation diagram from the ratio of column densities,

\begin{equation}
\frac{N_{j+2}}{N_{3}} = \frac{F_{j} \lambda_{j} A_{1}}{F_{1} \lambda_{1} A_{j}},
\label{eqn:obrat}
\end{equation}
\noindent where $N_{3}$ is the column density of the upper energy level of the transition S(1). We select $N_{3}$, corresponding to the S(1) transition, since it is the brightest and most frequently detected line. Using equation \ref{eqn:nuj}, we find this column density ratio from the continuous temperature model to be

\begin{equation}
\frac{N_{j+2}}{N_{3}} = \frac{\int_{T_{\ell}}^{T_{u}} \frac{g_{j+2}}{Z(T)} \times e^{\frac{-E_{j+2}}{kT}} T^{-n} dT}{\int_{T_{\ell}}^{T_{u}} \frac{g_{3}}{Z(T)} \times e^{\frac{-E_{1}}{kT}} T^{-n} dT}.
\label{eqn:mrat}
\end{equation}

We determined the parameters of our model by comparing the observed and the modeled column density ratios from equation \ref{eqn:obrat} and \ref{eqn:mrat},  by varying the lower temperature $T_{\ell}$ and power law index $n$ (the upper temperature, $T_{u}$, is fixed, see \S\,\ref{sec:method--calibration}).  Knowing these parameters, we substitute their values in equation \ref{eqn:ntot2}, to obtain the total column density. The total number of molecules is calculated using

\begin{equation}
n_{tot} = N_{tot} \Omega d^{2},
\end{equation} 
\noindent where $n_{tot}$ is the total number of hydrogen molecules and $d$ is the distance to the object. We then calculate the total $H_{2}$ gas mass,

\begin{equation}
M_{{H}_{2},tot} = n_{tot} \times m_{{H}_{2}},
\label{eqn:great}
\end{equation}
where $m_{H_2}$ is the mass of a hydrogen molecule.

\section{Method \& Calibration}
\label{sec:method--calibration}
Applying the model developed in \S\,\ref{sec:model} permits direct assessment of the warm \htwo\ column.  The model-derived lower temperature is typically degenerate, so extrapolating the model to lower temperatures to recover the full molecular content requires a calibration procedure using a trusted independent estimate of \htwo\ content. Here we describe both applications.

\subsection{Warm $H_{2}$}
The power law temperature distribution of $H_{2}$ molecules in Eq.\,\ref{eqn:powerlaw} has three primary parameters: upper temperature $T_u$, power law index $n$, and lower temperature $T_\ell$. An excitation diagram formed from the \htwo\ line flux ratios is the distribution of normalized level populations and constrains our model parameters. Excitation diagrams relate the column density of the upper level ($N_{u}$) of a particular transition, normalized by its statistical weight, $g_{u}$, as a function of its energy level $E_{u}$. In the following sections we describe how the model can be used to estimate the warm \htwo\ gas mass.  In practice, except in a few cases, the lower temperature cutoff $T_\ell$ is not directly constrained (see \S\,\ref{sec:determ-model-extr}).  The model fit results themselves are given in Table \ref{table:modelvalue}. 

\subsubsection{Upper temperature, $T_{u}$}
The bond dissociation energy for \htwo\ is 4.5 eV, corresponding to $E/k\sim5\times10^4$\,K, hence \htwo\ is not typically bound above this temperature. \htwo\ present in photo-dissociation regions (PDRs) in the outer layers of the molecular cloud can reach temperatures of a few 100 K \citep{Hollenbach97, Hollenbach99}. For the SINGS galaxies, \citet{Roussel07} concluded that the mass of \htwo\ greater than 100 K contributes 1--30\% of the total \htwo\ gas mass. In the MOlecular Hydrogen Emission Galaxies (MOHEGs) of \citet{Ogle10}, $\rm{M(\htwo)>1500}$ K contributes only 0.01$\%$ of the total \htwo.  $\rm{M(\htwo)>300}$ K contributes less than 1$\%$ in ULIRGs \citep{Higdon06}. 

In our model, when $T_{u}$ is allowed to vary above 1000\,K, we found negligible impact on the recovered total gas mass or the quality of the fit to the excitation diagram. We therefore fixed the upper temperature of the power law model distribution at $T_{u}=2000$ K. \htwo's ro-vibrational excitation spectrum (including the 1-0 S(1) line at $\lambda=2.12\micron$) would provide sensitivity to even hotter gas, but the mass contribution of this very hot gas in the context of our model is insubstantial.

\subsubsection{Power law index, n}
\begin{figure}
\centering
\includegraphics[width=0.50\textwidth]{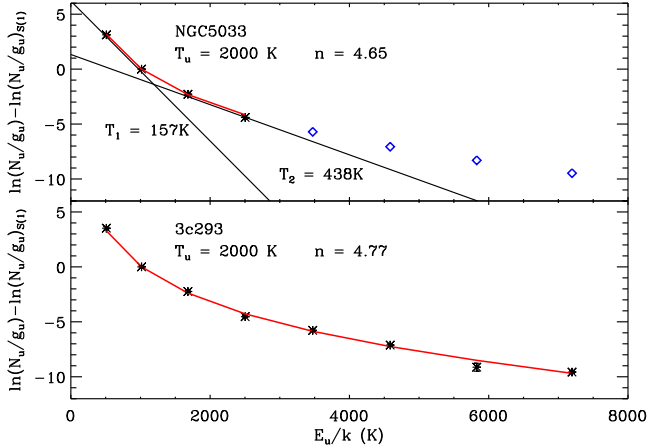}
\caption{Excitation Diagram for NGC\thinspace5033 and 3c\thinspace293. The $N_{u}$/$g_{u}$ ratios are normalized with respect to the S(1) transition. The solid red line indicates our model fit to the observed normalized columns, denoted by black points. The error bars on the black points are comparable to the symbol sizes.  The resulting model parameters $T_u$, and $n$ are indicated.  The two solid lines show the two discrete temperature fit adopted by \citet{Roussel07}. The blue diamonds show model-predicted ratio values for the (unobserved) S(4)--S(7) \htwo\ rotational lines in NGC\thinspace5033.  The lower panel for 3c\thinspace293 shows our model fit to all the MIR rotational lines S(0)--S(7) detected with the \emph{IRS-Spitzer}.  The radio jets in 3c293 shock-excite \htwo\ to high energy levels resulting in high rotational line fluxes.  A single power law model, varying only slope, reproduces \htwo\ excitation across a wide range of excitation energy. }
\label{fig:re}
\end{figure}

Keeping a fixed value of $T_{u}=2000$\,K, the other two model parameters, $T_\ell$ and $n$, are varied using a Levenberg-Marquardt optimization \citep{Markwardt09} to match the observed line flux ratios (suitably converted to column density ratios using Eq.~\ref{eqn:obrat}). Figure \ref{fig:re} is an example excitation diagram for galaxy NGC\thinspace5033. The model fit converges to $T_\ell=51$\,K, $n=4.65$, with a fixed value of $T_{u}=2000$\,K.  Model predicted ratio values for the unobserved lines S(4)--S(7) are indicated, as are the two discrete temperature fits chosen by \citet{Roussel07}. 

\begin{figure}
\centering
\includegraphics*[width=0.50\textwidth]{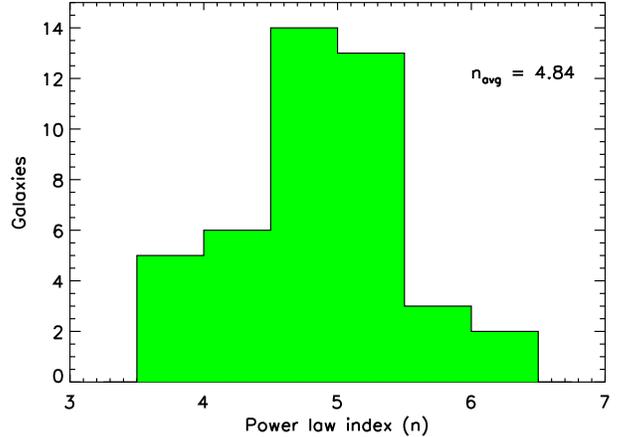}
\caption{The frequency distribution of the power law index, $n$, required to fit the MIR \htwo\ rotational lines. The average value $\bar{n}=4.84\pm0.61$ for the SINGS galaxy sample.}
\label{fig:fdspl}
\end{figure}

The power law index $n$ in our sample ranges from 3.79--6.39, with an average value $\bar{n}=4.84\pm0.61$.  Figure \ref{fig:fdspl} is a frequency distribution of power law index required to fit the MIR \htwo\ rotational line fluxes in the SINGS galaxies. The power law index range derived in our model is comparable to the indices required to fit the \htwo\ rotational line fluxes for ULIRGs (2.5 $<$ n $<$ 5.0) and Seyfert galaxies (3.4--4.9) \citep{Zakamska10, Pereira14}. 

Galaxies with a steep power law index have low warm gas mass fractions, since a larger quantity of \htwo\ in these systems are at lower temperatures.  Molecular gas heated by shocks and other turbulent energetic phenomena have higher temperatures and hence, higher warm gas mass fractions than gas in photo-dissociation regions (PDRs) (Appleton et al. in prep). The power law index therefore gives information on the relative importance of gas heating by shocks, photoelectric heating, UV pumping, etc.
 
\subsection{Total $H_{2}$}
With a continuous power law model well reproducing the rotational emission lines from warm \htwo, it is natural to consider whether the total molecular gas reservoir could be probed by suitable extrapolation of our model to lower temperatures. Typically, the entire reservoir of \htwo\ cannot be probed through rotational emission, since the model loses sensitivity at temperatures far below the first rotational energy state.  Recovering the total molecular content from rotational $H_{2}$ emission therefore typically requires an additional free parameter --- an \emph{extrapolated} model lower temperature, $T_\ell$ --- which must be calibrated against known molecular mass estimates.  We first explore the models sensitivity to low temperatures, and then describe the calibration procedure we adopt.

\subsubsection{Model Sensitivity Temperature, $T_s$} 
\label{sec:determ-model-extr}
The upper energy level of the lowest rotational transition of \htwo\ is 510\,K (see Table~\ref{table:lineppt}) --- substantially higher than typical kinetic temperatures in molecular regions.  Yet, even at temperatures well below this value, molecules in the high temperature tail of the energy distribution can yield detectable levels of rotational emission.  Below some limiting temperature, however, increasing the number of cold \htwo\ molecules will increase the implied molecular gas mass, but result in no measureable changes to the \htwo\ rotational line fluxes. 

\begin{figure}
\centering
\includegraphics*[width=0.5\textwidth]{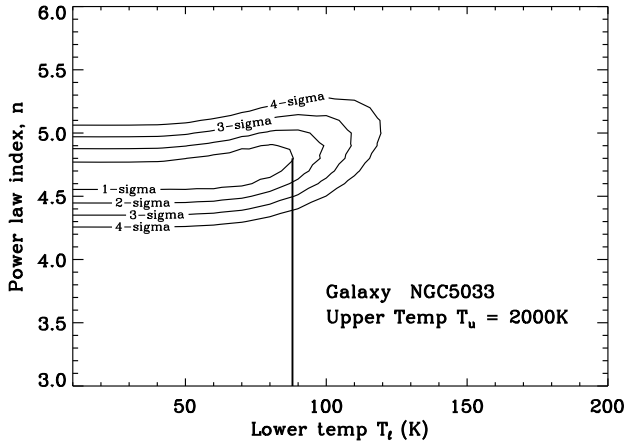}
\caption{The $\Delta \chi^{2}$ value in the model parameter space $T_{\ell}$ and n for NGC\thinspace5033. At temperatures lower than 88 K (the maximum $T_{\ell}$ value for 1$\sigma$ contour plot), the model yields similar flux ratios. Molecules with temperature less than  a sensitivity temperature $T_s=88$\,K have negligible contribution to \htwo\ rotational line flux in the galaxy NGC\thinspace5033.}
\label{fig:tisens}
\end{figure}

We define the sensitivity temperature, $T_{s}$, as that temperature below which the \htwo\ reservoir is too cold for changes in the amount of gas to lead to measurable changes in the excitation diagram. To evaluate this in the context of our model, we calculated the difference between the model-derived and observed column density ratios, $[N_{u}/g_{u}]/[(N_{u}/g_{u})_{S(1)}]$.  Note that with the adopted normalization to the $S(1)$ line, a unique excitation curve exists for any given pair $n$, $T_\ell$ ($\forall T_\ell \leq T_s$).  We varied these model parameters and evaluated the quality of the model fit using
\begin{equation}
\chi^{2} = \sum_{i=1}^m \left(\frac{R_{i,mod} - R_{i,obs}}{\sigma_{R_{i,obs}}}\right)^2,
\end{equation}
where $R = \ln\left(\frac{N_{u}/g_{u}}{(N_{u}/g_{u})_{S(1)}}\right)$, and $R_{i,mod}$ and $R_{i,obs}$ are, respectively, the modeled and observed flux ratios for the $i^{th}$ transition, with uncertainty $\sigma_{R_{i,obs}}$, and the summation is over all independent line flux ratios.  We map the $\chi^2$ space in $T_\ell$ and $n$ to $\sigma$ values, following \citet{Avni76}, via $\Delta\chi^{2}=\chi^{2} - \chi_{min}^{2}$ (where $\chi_{min}^{2}$ is the minimum $\chi^{2}$ value).  Figure \ref{fig:tisens} shows example $\Delta\chi^{2}$ contours for the galaxy NGC\thinspace5033. The value of $T_s$ is determined at the maximum value of lower cutoff temperature $T_\ell$ along the $1\sigma$ contour. For the galaxy NGC\thinspace5033, $T_{s}\sim88$\,K.

The near-horizontal orientation of the contours indicates little correlation between $n$ and $T_\ell$.  Since for most of the sample, the contours do not close as you go towards lower temperatures $T_\ell<T_s$, any chosen lower cutoff of the power law distribution of temperatures below $T_{s}$ remains consistent with the data.\footnote{This indicates that the precise form of the temperature distribution at $T<T_s$ is not well constrained; see \S\,\ref{sec:broken-power-law}.}

\begin{figure}
\centering
\includegraphics*[width=0.5\textwidth]{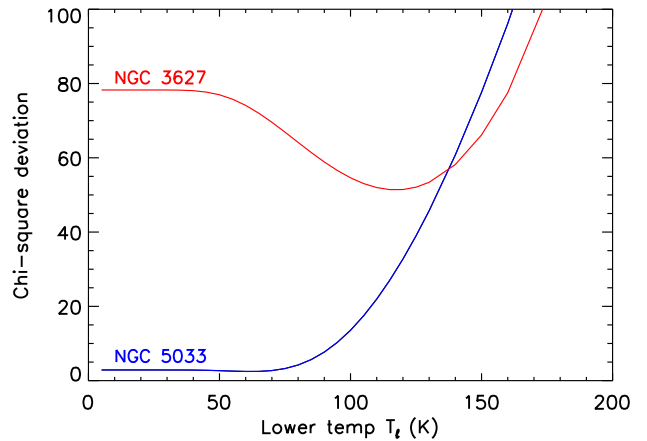}
\caption{The $\chi^{2}$ distribution for a normal galaxy NGC\thinspace5033 (blue) and an excess warm molecular gas galaxy NGC\thinspace3627 (red). The minimum $\chi^{2}$ occurs at a cutoff temperature of 120 K for NGC\thinspace3627 however, for the galaxy NGC\thinspace5033 the $\chi^{2}$ curve decreases throughout as the lower temperature is decreased. For temperatures less than 50 K the model show no change in the flux ratios since temperatures are too cold to occupy rotational energy levels and hence, a constant $\chi^{2}$ curve.}
\label{fig:norm_warm}
\end{figure}

Generally, differences between the modeled and observed flux ratios decrease as $T_\ell$ decreases, and do not significantly change below $\sim$80 K.  However, in some LINER and Seyferts systems (e.g. NGC\thinspace2798, NGC\thinspace3627, and NGC\thinspace4579), the contours \textit{close}, and $\chi^{2}$ has a defined minimum. Figure \ref{fig:norm_warm} illustrates this phenomenon in the galaxy NGC\thinspace3627, evaluating $\chi^2$ along the ridge line of best-fitting indices $n$ for each $T_\ell$, and contrasting this galaxy with the more typical case of an indefinite minimum.  Evident in the fits to NGC\thinspace3627 is a single best-fitting lower cutoff temperature of $T_\ell\sim120$\,K.  In such cases, a continuation of the power-law distribution to temperatures below this best-fitting value is counter-indicated, as it degrades the fit.  This indicates that in these cases, the bulk of \htwo\ is directly detected via rotational emission in a warm gas component excess. 

\begin{figure}
\centering
\includegraphics*[width=0.5\textwidth]{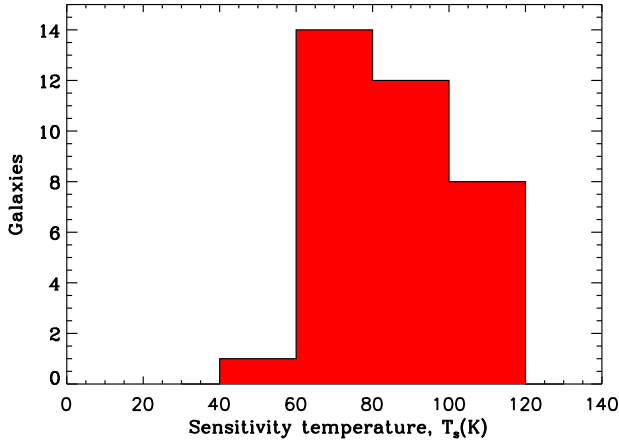}
\caption{The distribution of sensitivity temperatures $T_{s}$ (the temperature above which \htwo\ can be directly traced via rotational emission) for the SINGS sample, omitting warm excess sources. On average, the sensitivity temperature is 81\,K.}
\label{fig:fdts}
\end{figure}

\begin{figure}
\centering
\includegraphics*[width=0.5\textwidth]{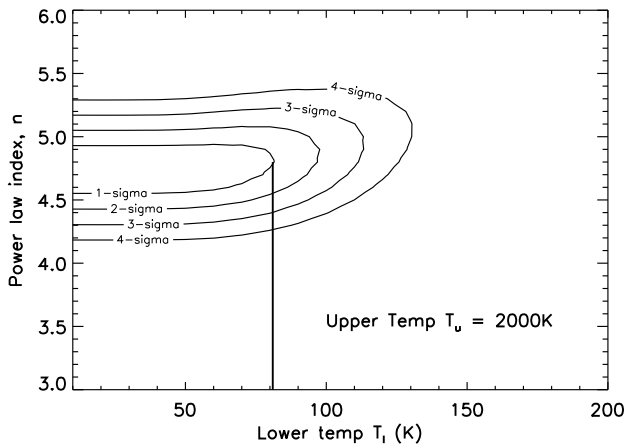}
\caption{The average $\Delta \chi^2$ contours of the SINGS sample, excluding warm excess galaxies. The average value is $\overline{T_s}=81$\,K with an average power law index of $\overline{n}=4.8$. Using these parameters, $M_{H_2}(>81K)/M_{H_2,total}=15\%$ of the \htwo\ mass is probed directly via rotational emission (\S\,\ref{sec:mass-distr-funct}). }
\label{fig:totcont}
\end{figure}

To identify galaxies with warm molecular gas excess, we imposed the constraint $\Delta\chi^{2} =  \chi_{20K}^{2} - \chi_{min}^{2} > 2.3$, where $\chi^{2}_{20 K}$ is the $\chi^{2}$ value at 20 K. Figures \ref{fig:fdts} and \ref{fig:totcont} show the distribution of sensitivity temperature $T_{s}$ and the corresponding average confidence contour plot for all SINGS galaxies, excluding in total 8 galaxies, identified as having a warm gas excess (NGC\thinspace2798, NGC\thinspace3627, and NGC\thinspace4579) and few other galaxies (e.g. NGC\thinspace1266, NGC\thinspace1291, NGC\thinspace1316, NGC\thinspace4125, and NGC\thinspace5195) with low signal-to-noise (S/N) ratio in their S(0) line. The excluded galaxies are all LINERs or Seyferts with low S(0)/S(1) ratios (in the range 0.04--0.2, vs. median 0.33 in the SINGS sample, \citet{Roussel07}).  They exhibit evidence for shocks, unusually high dust temperatures, and merger morphologies --- all processes that could result in substantial quantities of warm molecular gas.


As seen in Fig.~\ref{fig:totcont}, the average value of $T_{s}$ is 81\,K.  This implies that, with the quality of \htwo\ rotational spectroscopy provided by the Spitzer/IRS instrument, \htwo\ gas down to rotational temperatures of $\sim$80\,K can be reliably detected.

\subsubsection{Model extrapolated lower temperature, $T_{\ell}$}
\label{sec:model-extr-lower}
Given our model sensitivity to \htwo\ columns down to rotational temperatures $\sim$80\,K, we evaluate the potential for \textit{extrapolating} the power law distribution to lower temperatures to attempt to recover the the total molecular gas mass. 

To calibrate an extrapolated lower temperature cutoff --  $T_{\ell}$ ---  we require a set of galaxies with well-established molecular gas masses obtained from $\rm{L_{CO}}$.  We constructed a training sample from SINGS galaxies (except NGC\thinspace4725, where we have an upper mass limit of molecular gas from CO intensity), omitting galaxies with a central warm excess (see \S\,\ref{sec:determ-model-extr}).  ULIRGs and radio galaxies are also avoided due to ambiguity in their $\rm{\alpha_{CO}}$ values. 

Recent studies have also shown $\rm{\alpha_{CO}}$ in low metallicity galaxies can be substantially higher than the Galactic value \citep[e.g.][]{Bolatto13, Schruba12, Leroy11} and hence we restricted our sample to galaxies with an oxygen abundance value of $12+\log[\mathrm{O}/\mathrm{H}]\geq8.4$, using the average of their characteristic metallicities derived from a theoretical calibration and an empirical calibration as recommended by \citet{Moustakas10}. 

To perform the calibration, the single power law model fitted to the rotational excitation diagrams of each of the 34 galaxy training sample is extrapolated to the temperature where the model mass equals the cold molecular gas mass, estimated from CO emission ($\rm{L_{CO}}$), using $\rm{\alpha_{CO,Gal}}$ (column 10 of Table \ref{table:h2flux}). 

\begin{figure}
\centering
\includegraphics*[width=0.50\textwidth]{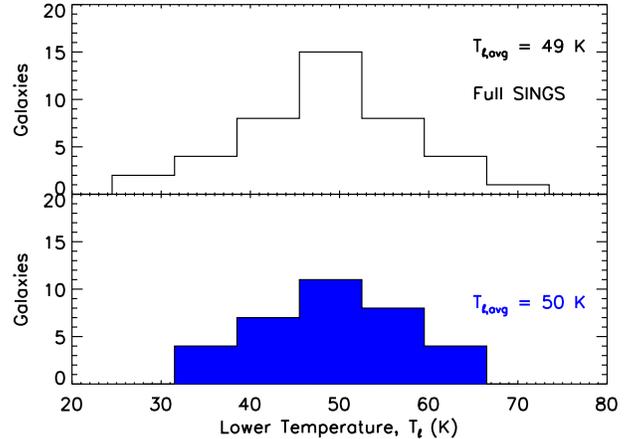}
\caption{The frequency distribution of model lower temperature, $T_{\ell}$, required to fit the molecular gas mass using the Galactic $\rm{\alpha_{CO,Gal}}$. On average the value for $T_{\ell}$ is $T_\ell^\star=$49 K for our training sample. The below histogram (in blue) is plotted excluding warm galaxies with high $T_{s}$. However, no change in the average value, $T_{\ell}^\star$, is found including or excluding galaxies with warm gas excess.}
\label{fig:fds}
\end{figure}

Figure \ref{fig:fds} shows the distribution of extrapolated model lower temperature, $T_{\ell}$, required to fit the molecular gas mass in the SINGS sample (with and without warm excess sources). The full training sample is described by an average extrapolation temperature of $T_{\ell}^{\star} = 49\pm9$\,K. Excluding galaxies with warm gas excess narrows the distribution, but does not significantly change the average value of $T_{\ell}$.  Table \ref{table:modelvalue} lists the value of $T_{\ell}$ for each galaxy, with the value in parentheses calculated when the conversion factor derived by \citet{Sandstrom13} through dust emission is assumed. 

It is important to note that the numerical value of the lower temperature cutoff of $T_{\ell}^{\star} = 49\pm9$\,K relies on the assumption of a single power law across all temperatures. See \S\,\ref{sec:broken-power-law} for an alternative viewpoint adopting broken power laws, which can change the value of $T_\ell^\star$ without impacting the estimated total molecular gas mass.

\subsubsection{Mass distribution function} 
\label{sec:mass-distr-funct}
Since $\rm{dM/dT}$ is a strong negative power law, the number of molecules decreases rapidly with increasing temperature --- most of the \htwo\ gas mass resides at the lowest temperatures. We can evaluate the fraction of the total molecular mass our model is sensitive to by comparing $\rm{M(T=T_\ell\rightarrow T_s)}$ and $\rm{M(T=T_s\rightarrow T_u)}$

\begin{figure}
\centering
\includegraphics*[width=0.50\textwidth]{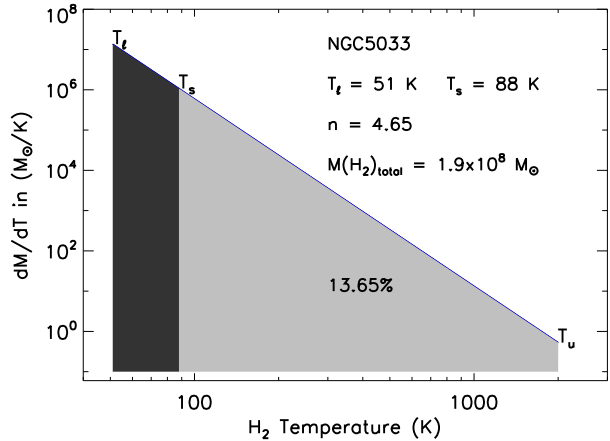}
\caption{The distribution $dM/dT$ vs molecular temperature for galaxy NGC\thinspace5033.  The darker and lighter shades in the plot show two different regions, below and above the sensitivity temperature, $T_s$ = 88 K. The fraction of \htwo\ gas mass below $T_s=88\,\mathrm{K}$ for the galaxy NGC\thinspace5033 is about $86\%$.}
\label{fig:dmt}
\end{figure}
Figure \ref{fig:dmt} shows an example mass-temperature distribution $dM/dT$ for NGC\thinspace5033 as a function of molecular temperature. The total molecular gas mass of 1.9 $\times 10^{8}$ M$_\odot$ is distributed with an index $n=4.65$ in the temperature range $T_\ell=51$\,K to $T_u=2000$\,K.  The fraction of cold \htwo\ gas mass ($<T_{S}=88$\,K for NGC\thinspace5033) is $\sim86\%$.

Using the average power-law index and sensitivity temperature from the SINGS training sample (see Fig.\,\ref{fig:totcont}) we find $M_{H_2}(T>T_s=81\mathrm{K})/M_{H_2,\mathrm{total}}\sim15\%$. Adopting instead the 2$\sigma$ contour, the value of $T_s$ becomes 97\,K and the ratio M$_{H_{2}}(>T_s=97\mathrm{K})/M_{H_{2},\mathrm{total}}$ is $\sim8\%$. Even when considering a broken power law temperature distribution for H$_{2}$, the mass fraction remains similar, and we discuss the effects in \S\ref{sec:effect-dust-temp}. This demonstrates that, despite the many shortcoming of \htwo\ as a rotational emitter, \emph{we can trace a substantial fraction of the \htwo\ in galaxies directly.}

\section{Results, Discussions, \& Applications}

\subsection{What are the typical molecular gas temperatures in galaxies?}
UV pumping sets the level populations of upper level states, J $\geqslant$ 3, in particular at low column densities of \htwo. As a result, these states may result in rotational temperatures well in excess of their kinetic temperatures. For our purposes, whatever combination of collisional (including shocks) and UV pumping determines the level populations and hence average the \htwo\ excitation, we assume a single power law distribution of rotational temperature describes the ensemble of molecules (although see \S\,\ref{sec:broken-power-law} for an alternative analysis employing broken power-laws). Since the mass is dominated by gas with the lowest rotational temperatures and hence lowest excitations, it is instructive to compare the derived lower temperature cutoffs (\S 4.2.2) with various theoretical models and measurements of temperature in molecular gas.

The average lower extrapolation cutoff, $T_\ell^\star=49$\,K, is comparatively higher than what is typically assumed for molecular regions ($\sim$10--30\,K).  And yet, this high lower cutoff is essential in describing the portion of molecular material following a single power-law distribution of temperatures.  For example, if we forced the extrapolation temperature to a lower value of 20\,K while retaining the average power law index $n=4.8$ (\S\,4.1.2), the estimated molecular gas mass would be on average $\sim30\times$ higher than the mass measured using $\rm{L_{CO}}$. 

The ratio of warm diffuse to cold dense molecular gas in galaxies is a key step in understanding the \htwo\ temperature distribution. Assuming an incident radiation field G$_{0}$ = 10 for a cloud with mass 10$^{5}$--10$^{6}$ M$_{\odot}$, \citet{Wolfire10} estimated that molecules in the outer diffuse region ($A_{V} <1$) obtain temperatures in the range 50--80\,K, with stronger incident radiation fields leading to even higher temperatures.  Due to the differences in self shielding from dissociating radiation, molecular clouds have an outer layer of varying thickness which contains molecular hydrogen, but little or no CO. This gas cannot be traced through CO transitions and is hence called \emph{dark molecular gas} \citep{Wolfire10}. This dark gas can account for a significant fraction (24--55\%) of the total molecular gas mass in galaxies \citep{Pineda13, Wolfire10, Smith14, Planck11}.

CO-dark and diffuse molecular gas is heated to higher temperatures than dense CO-emitting gas in  molecular cloud cores.  Indeed, individual molecular cloud simulations find average mass-weighted \htwo\ temperatures of $\sim$45\,K for Milky-Way average cloud masses and radiation (\citet{Glover12a}, Glover, priv comm.). Galaxy-scale hydrodynamic simulations with a full molecular chemistry network \citep{Smith14} also show that \htwo\ gas in the CO-emitting regions is markedly biased to the low temperature end of the full temperature distribution (Glover, priv comm.). Also, in diffuse and translucent molecular clouds in the Galaxy, \citet{Ingalls11} demonstrated that additional sources of heating are required to explain the observed \htwo\ and atomic cooling-line power. Our extrapolated power law model traces both warm and cold molecular gas mass in galaxies.  The common assumption that the bulk of molecular gas is found at rotational temperatures of $\sim$10--30 K may be true only in the CO emitting core regions of the molecular cloud.  The average mass-weighted molecular gas temperature can be much higher when the full range of emitting environments are considered.  It is therefore perhaps not surprising to find typical molecular gas temperatures of $\sim50$\,K in galaxies, similar to our model derived values. 

It must be noted, however, that to calculate the total molecular gas mass, we have extrapolated using a single power law index. The possibility of a broken or non-power-law temperature distribution for the $\sim$85\% of \htwo\ gas at temperatures lower than our sensitivity temperature cannot be excluded (and indeed in warm gas excess sources, will be required to explain ongoing star formation).  In the following section, we consider the effect of broken power law models on the dominant \htwo\ temperatures.

\subsection{Broken power law models}
\label{sec:broken-power-law}
In the previous subsection we discussed the results of applying a single power law model for the entire temperature distribution of \htwo, and found that a single lower cutoff temperature of $T_{\ell}^{\star}$ = 49\,K permits recovering the entire molecular mass.  Yet as seen in \S\,\ref{sec:determ-model-extr}, with the typical performance of the Spitzer/IRS instrument, we are sensitive only to the $\sim$15\% of \htwo\ gas with temperature $T\gtrsim T_s=81$\,K.  While a single power law temperature distribution, varying only the slope $n$, provides very high quality model fits to the \htwo\ excitation diagram, at temperatures below $T_s$, in principle \emph{any} distribution form would provide an equally valid description.  Chemo-hydrodynamic models with \htwo\ formation and excitation do result in a broad distribution of \htwo\ temperatures (Glover \& Smith, in prep.).  But shocks elevate \htwo\ to higher rotational temperatures \citep[e.g.][]{Neufeld08}, and are not well modeled, so as yet there exists no \emph{a priori} expectation for the detailed 10--3000\,K temperature distribution of \htwo\ molecules.  Any presumed distribution, however, must 1) reproduce the \htwo\ excitation diagrams as well as a single power law does, and 2) permit recovering the CO-predicted \htwo\ mass within a reasonable range of temperatures.  A \emph{broken power-law} provides a general framework for exploring changes in the low-temperature form of the distribution, and has been invoked by \citet{Pereira14} to jointly model \htwo\ and CO excitation.  Here we consider the changes resulting from adopting a broken power law model, with a fixed break temperature,  $T_{b}$, and slope below the break, $n_1$, studying the resulting lower cutoff temperatures, as well as the ability of these models to satisfy these two criteria.  

Holding the upper temperature $T_u$ constant as before, we evaluated a grid of broken power law models with fixed pre-break slope $n_1$ (from 0--3) and break temperature $T_{b}$ (from 70--290\,K). For each such model, we fitted the observed MIR \htwo\ rotational line ratios in excitation diagrams in the same training sample of 34 galaxies considered in \S\,\ref{sec:model-extr-lower}. As before, we then extrapolate the broken power law to recover the total \htwo\ mass, as indicated by CO intensity, evaluating the associated extrapolation cutoff temperature $T_{\ell,b}$.  We define $T_{\ell,b}^{\star}$ as the average value of $T_{\ell,b}$ across the training sample, analogous to $T_{\ell}^{\star}$ for a single power law (\S\,\ref{sec:model-extr-lower}).

\begin{figure}
\centering
\includegraphics*[width=0.5\textwidth]{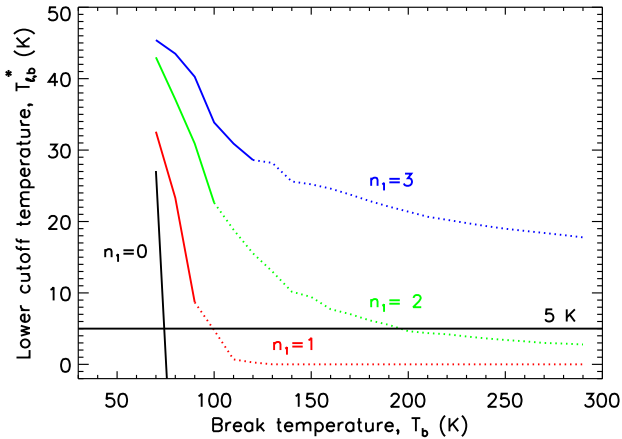}
\caption{The lower extrapolation temperature for broken power-law models, $T_{\ell,b}^{\star}$, as a function of the model break temperature, $T_{b}$ for different values of lower power law indices $n_1$ = 0 (black), 1 (red), 2 (green), and 3 (blue). The solid horizontal line denotes a lower temperature physical cut off of 5\,K. The dotted line in the figure shows the range where the model poorly fits the MIR $\htwo$ rotational lines, with median $\Delta\chi^{2} > 2.3$. At break temperatures above $\sim$120\,K, for all n$_{1}$ (0--3), the $\htwo$ rotational lines are poorly fit.}
\label{fig:fig9}
\end{figure}

Figure \ref{fig:fig9} shows the variation of the training sample's average extrapolated lower cutoff temperature $T_{\ell,b}^{\star}$ as a function of break temperature $T_{b}$ for differing values of below-the-break slope. Those combinations of ($n_1$,$T_b$) for which the excitation diagrams are reproduced as well as a single power law model ($\Delta\chi^{2} < 2.3$ relative to the single power law model fit; see \S\,\ref{sec:determ-model-extr}) for at least 50\% of the training sample are shown in solid lines.  In this allowable range of $n_1$ and $T_b$, the dispersion of the ratio $M(\htwo,\mathrm{model})/M(\htwo,\mathrm{CO})$ of \htwo\ mass recovered from the broken power-law model using a single extrapolation $T_{\ell,b}^{\star}$, relative to that implied by CO ranges only from 0.28--0.34 dex.  At higher temperatures the excitation diagram fit quality degrades rapidly (dotted lines).  As the break temperature approaches the average slope for a single power law model, $T_b=T_\ell^\star=49$\,K, the broken power law becomes equivalent to our single power law truncated at this temperature, independent of slope $n_1$.  At slopes approaching the single-power law average $n=4.8$ (see Fig.~\ref{fig:fdspl}), lower extrapolated temperatures diverge less and less from the value obtained using a single power law, since the shape of the temperature distribution is then relatively unchanged.  Low slopes $n_1$ require very low extrapolation temperatures, since the cumulative mass of \htwo\ rises less rapidly at lower temperatures compared to the single power law.  In fact, models with a flat slope ($n_1=0$) fail to recover the CO-derived \htwo\ mass within a reasonable lower \htwo\ temperature floor of $\sim5$\,K.

Figure \ref{fig:fig9} demonstrates that the single lower cutoff temperature to which an \htwo\ model must be extrapolated depends sensitively on the detailed shape of the distribution at low temperature.  For reasonable slopes, the dominant mass-weighted temperature can range from 10--50\,K.  All models which produce satisfactory fits of the excitation diagrams and have physically reasonable extrapolation temperatures have nearly the same predictive capability for the \emph{total} \htwo\ mass. While optical depth effects introduce some difficulties (e.g. for low-J $^{12}$CO), ideally a family of profiles for the full temperature distribution of molecular gas in galaxies could be developed, based on tracers with sensitivity to different parts of the temperature range matched to chemo-hydrodynamic models of molecular gas across all densities. Since the single power law model provides excellent fits to \htwo\ rotational emission line ratios with the minimum of free parameters, we retain this simplified model for the remainder of this work.  It should, however, be kept in mind that broken power law models with flatter slopes below $T\sim$100--120\,K can produce similar quality fits and yield equivalently reliable model-based \htwo\ masses, as long as the low temperature slope $n_1$ does not vary substantially from galaxy to galaxy.  

\subsection{Estimating total $M_{H_{2}}$}

\begin{figure*}
\centering
\includegraphics*[width=1.00\textwidth]{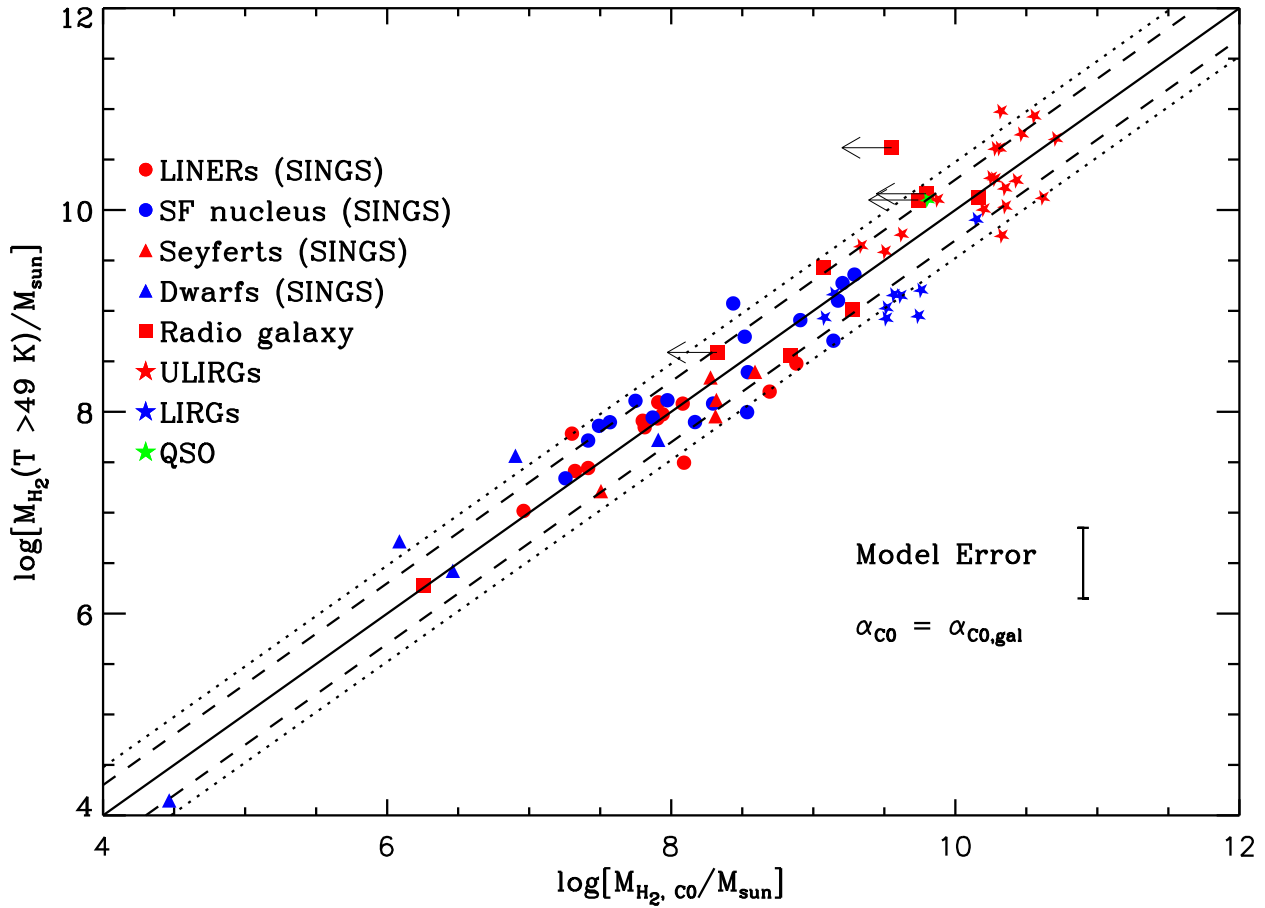}
\caption{The model extrapolated molecular gas mass, assuming a single lower cutoff temperature of 49\,K vs the total molecular gas mass obtained from L$\rm{_{CO}}$ measurements using $\rm{\alpha_{CO,Gal}}$. Different symbols represent different galactic systems as indicated. The solid line is one to one correspondence while the dashed and dotted lines are $2\times$ and $3\times$ the value respectively.}
\label{fig:49co}
\end{figure*}

By fitting a continuous temperature distribution to the MIR \htwo\ rotational lines and calibrating an extrapolating temperature, this model can be used to calculate the total molecular mass in galaxies directly from \htwo\ rotational emission. While this method does require a reliable source of known molecular masses for calibration (introducing secondary dependence on, e.g. $\alpha_{CO}$), once calibrated it is independent of any indirect tracer like CO, DGR, or assumptions about star formation depletion timescales.  The biases inherent in this method (arising, for example, from the assumption of a single and smooth power law temperature distribution), are therefore expected to be relatively distinct from those of the aforementioned estimators. In this section we test the model's capability to estimate molecular gas mass in different types of galaxies.

Adopting a fixed model lower extrapolation temperature $T_{l}^{\star} = 49$ K the total \htwo\ gas mass is calculated by extrapolating the fitted model. The total \htwo\ gas mass derived by our model is shown in Figure \ref{fig:49co}. It compares very well with $\rm{L_{CO}}$ along with $\rm{\alpha_{CO,Gal}}$ based estimates of gas mass. The scatter in our model is about 0.31 dex (factor of 2) for the SINGS sample and increases to 0.34 dex (factor of 2.2) for the complete sample, including U/LIRGs and radio galaxies. Some of this scatter no doubt arises from ignorance of the true $\rm{\alpha_{CO}}$ values in these systems. 

Galaxies with warm gas excess and a high sensitivity temperature $T_{s}$ (\S\,\ref{sec:determ-model-extr}), require similar values of extrapolated $T_{\ell}$ as the training sample. The warm molecular gas traced by MIR rotational lines may be completely isolated from the cooler gas in these galaxies.  Extrapolating to a lower temperature of 49 K yields molecular gas masses which agree with CO-derived values, assuming $\rm{\alpha_{CO}=\alpha_{CO,Gal}}$.  A possibility of enhanced CO excitation due to high molecular gas temperature with a corresponding low $\rm{\alpha_{CO}}$ in such warm galaxies cannot be ruled out (see also \S\,\ref{sec:model-deriv-molec}). 

Taken together, a tight correlation between \htwo-derived and CO-based mass estimates is found, spanning seven orders of magnitude in mass scale and across a wide range of galaxy types.  The mass calculated with a continuous temperature distribution model, extrapolated to a single \emph{fixed} cutoff temperature of 49\,K can provide an independent measurement of total molecular gas mass in galaxies, good to within a factor of 2.2 (0.34 dex).  A dependence on $\alpha_{\mathrm{CO}}$ does arise indirectly through our CO-based mass estimates in the training sample (\S\,\ref{sec:model-extr-lower}), but the recovered dispersion is comparable to uncertainties in the $\rm{\alpha_{CO}}$ conversion factor itself, even among normal galaxies \citep{Bolatto13}, as well as methods using DGR \citep[e.g.][]{Sandstrom13}.

\subsection{Model derived molecular gas mass in ULIRGs, LIRGs and radio galaxies}
\label{sec:model-deriv-molec}
Many local ULIRGs are recent ongoing galaxy mergers. In the merging process a large amount of gas in the spiral disk is driven to the central nuclear region, increasing the gas temperature. The increase in temperature and turbulence increases the CO linewidth, resulting in a high value of $\rm{L_{CO}}$ for a given molecular gas mass.  $\rm{\alpha_{CO,Gal}}$ therefore gives an overestimate of \htwo\ gas masses in these galaxies \citep{Downes93, Bryant99}. Moreover, $\rm{\alpha_{CO,Gal}}$ can yield molecular gas masses greater than the observed dynamical mass \citep{Solomon97}. To avoid this, a lower value of conversion factor is suggested for ULIRGs and other merger systems --- $\rm{\alpha_{CO}}$ = 0.8 M$_\odot$(K km s$^{-1}$ pc$^{2})^{-1}$, 5.5$\times$ lower than the standard Galactic value \citep{Downes98}. However, by considering the high-J CO ladder, some studies have suggested that even for ULIRGs $\rm{\alpha_{CO,Gal}}$ values are possible \citep{Papadopoulos12}.  Some \htwo\ emission may lie outside photo-dissociation and star forming regions in ULIRGs \citep{Zakamska10}. This \htwo\ may reside in CO-dark gas, so that applying a low $\rm{\alpha_{CO}}$ value in ULIRGs may yield an underestimate \htwo\ mass.  

In radio galaxies molecular gas can be predominantly heated by shocks through powerful jets. The molecular gas clouds may be affected by turbulence, and not gravitationally bound, defying the use of standard $\rm{\alpha_{CO,Gal}}$. \citet{Ogle14} using DGR in NGC\thinspace4258, a low luminosity AGN (LLAGN) harboring a jet along the disk, derived gas mass of about 10$^{8}$ M$_\odot$, an order of magnitude lower compared to the standard method of using $\rm{\alpha_{CO,Gal}}$. The molecular gas mass could be overestimated when used $\rm{\alpha_{CO,Gal}}$ in radio galaxies, which harbor long collimated powerful jets.

In applying a power law model to the sample of ULIRGs, LIRGs and radio galaxies, the nominal model extrapolation temperature $T_\ell^\star=49$\,K is used to calculate the total \htwo\ gas masses, listed in column 3 of Table \ref{table:uli}. The $T_{\ell}$ in column 4 of Table \ref{table:uli} is the required extrapolated temperature to match the cold molecular gas mass measured from the CO line intensity.

In radio galaxies 3c\thinspace424, 3c\thinspace433, cen\thinspace A, and 3c\thinspace236 the estimated \htwo\ gas mass using the power law model after extrapolation to $T_{\ell}^{\star}$ = 49 K, is higher when compared to the CO luminosity derived values using $\rm{\alpha_{CO,Gal}}$. However, when accounted for the intrinsic variation in $\rm{\alpha_{CO,Gal}}$ and $T_{\ell}^{\star}$ the \htwo\ gas masses are in agreement with each other except in 3c\thinspace424, where the difference in masses is more than 10$\times$.

Using $\rm{\alpha_{CO}}$ = 0.8 M$_\odot$(K km s$^{-1}$ pc$^{2})^{-1}$, a factor of 5.5$\times$ lower than the standard Galactic value, we derive a modified extrapolation temperature $T'_{\ell}$, which is required to match the lowered gas mass. Since the model mass rises rapidly to lower temperatures, $T'_{\ell} > T_{\ell}$.  The $T'_{\ell}$ values are listed in column 7 of Table \ref{table:uli}. On average for ULIRGs, $T'_{\ell}=80\pm13$\,K. 

\begin{figure}
\centering
\includegraphics*[width=0.5\textwidth]{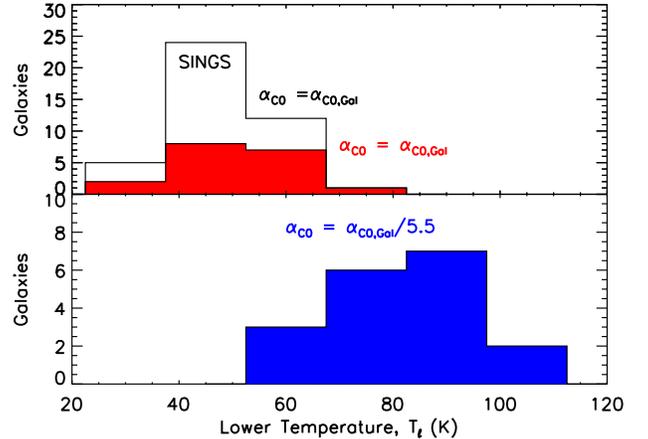}
\caption{The distribution of model extrapolated lower temperatures in ULIRG and radio galaxies when the \htwo\ gas mass is evaluated using the Galactic conversion factor $\rm{\alpha_{CO,Gal}}$ (above, red), and when  adopting $\rm{\alpha_{CO,Gal}/5.5}$, as generally accepted for ULIRGs (below, blue). The mean lower temperature cutoff is 50\,K and 80\,K when $\rm{\alpha_{CO,Gal}}$ and $\rm{\alpha_{CO}}=\rm{\alpha_{CO,Gal}}/5.5$ are used, respectively.  The $T_\ell$ distribution for the SINGS normal galaxy sample is shown above, for comparison.}
\label{fig:ulihist}
\end{figure}

Figure~\ref{fig:ulihist} shows the temperature distribution of $T_{\ell}$ (adopting $\rm{\alpha_{CO,Gal}}$) and $T'_{\ell}$ (adopting $\rm{\alpha_{CO,Gal}}/5.5$) in ULIRGs.  The lower temperature cutoff in ULIRGs and radio galaxies (except 3c424) is very similar to the normal galaxy training sample when $\rm{\alpha_{CO,Gal}}$ is adopted, but much higher with the reduced molecular mass of $\alpha_\textrm{CO,ULIRG}$.   

It is of interest that when $\alpha_\textrm{CO,Gal}$ is used with the nominal calibration cutoff temperature $T_\ell^\star$, the ULIRG sample in Fig.~\ref{fig:49co} does not exhibit any particular bias.  Either $\alpha_\textrm{CO}$ and $T_\ell$ are similar to their normal Galactic values in these systems, or they have reduced $\alpha_\textrm{CO}$ and a higher \htwo\ temperature floor. This may indicate that the same physical processes that lead to reduced $\alpha_\textrm{CO}$ in highly active system, including increased ISM pressure and radiation density, globally elevate the gas temperature. 

Since ULIRGs could indeed have uniformly elevated molecular gas temperatures, this degeneracy between $\rm{\alpha_{CO}}$ decrease and $T_\ell$ increase leads to a systematic uncertainty in the total \htwo\ gas mass identical in form to that obtained directly from mass estimates based on $\rm{L_{CO}}$.  A suggested prescription which side-steps this ambiguity, in applying this model to systems with non-Galactic $\rm{\alpha_{CO}}$ is to calculate the total gas mass using the the nominal $T_\ell^\star=49$\,K, and scale it by $\alpha_\textrm{CO}/\alpha_\textrm{CO,Gal}$, for the preferred $\alpha_\textrm{CO}$.  As can be seen in Fig.~\ref{fig:49co}, which adopts a uniform $\rm{\alpha_{CO,Gal}}$, the molecular mass in ULIRGs is well recovered by this procedure.

\subsection{Effect of dust temperature on the warm \htwo\ fraction}
\label{sec:effect-dust-temp}
Since typical sensitivity temperatures are $T_s\sim80$\,K (see \S\,\ref{sec:determ-model-extr}), we can directly calculate the warm \htwo\ mass above $\sim100$\,K without extrapoloation, and compare it to the total mass as recovered by the calibrated extrapolation of \S\,\ref{sec:model-extr-lower}.  Given an estimate for the power law index, $n$, and lower cut off temperature, $T_{\ell}$, of the power law distribution, we can calculate the fraction of molecular gas mass at temperatures above 100\,K as
\begin{equation}
\frac{M(>100\,\mathrm{K})}{M_{\mathrm{total}}} = \frac{M(>100\,\mathrm{K})}{M(H_{2},CO)} = \frac{\int_{100\mathrm{K}}^{T_{u}} T^{-n} dT}{\int_{T_{\ell}}^{T_{u}} T^{-n} dT},
\end{equation}
where M$_{\mathrm{total}}$ is the total molecular gas mass estimated from the CO line intensity. Assuming $100\,\mathrm{K} \ll T_{u}$ and $T_{\ell} \ll T_{u}$ we find
\begin{equation}
\frac{M(>100\,\mathrm{K})}{M_{\mathrm{total}}} \approx \left(\frac{100\,\mathrm{K}}{T_{\ell}}\right)^{1-n}.
\label{eqn:warmfrac}
\end{equation}

Table \ref{table:modelvalue} lists the calculated mass fraction $\rm{M(> 100 K)/M_{total}}$ for each galaxy. Columns 6 and 8 of Table \ref{table:uli} are calculated using $\rm{\alpha_{CO,Gal}}$ and $\rm{(1/5.5)\times\alpha_{CO,Gal}}$ as generally assumed for ULIRGs, respectively.  At lower $\rm{\alpha_{CO}}$, the warm gas mass fraction is higher.

\begin{figure*}
\centering
\includegraphics*[width=1.0\textwidth]{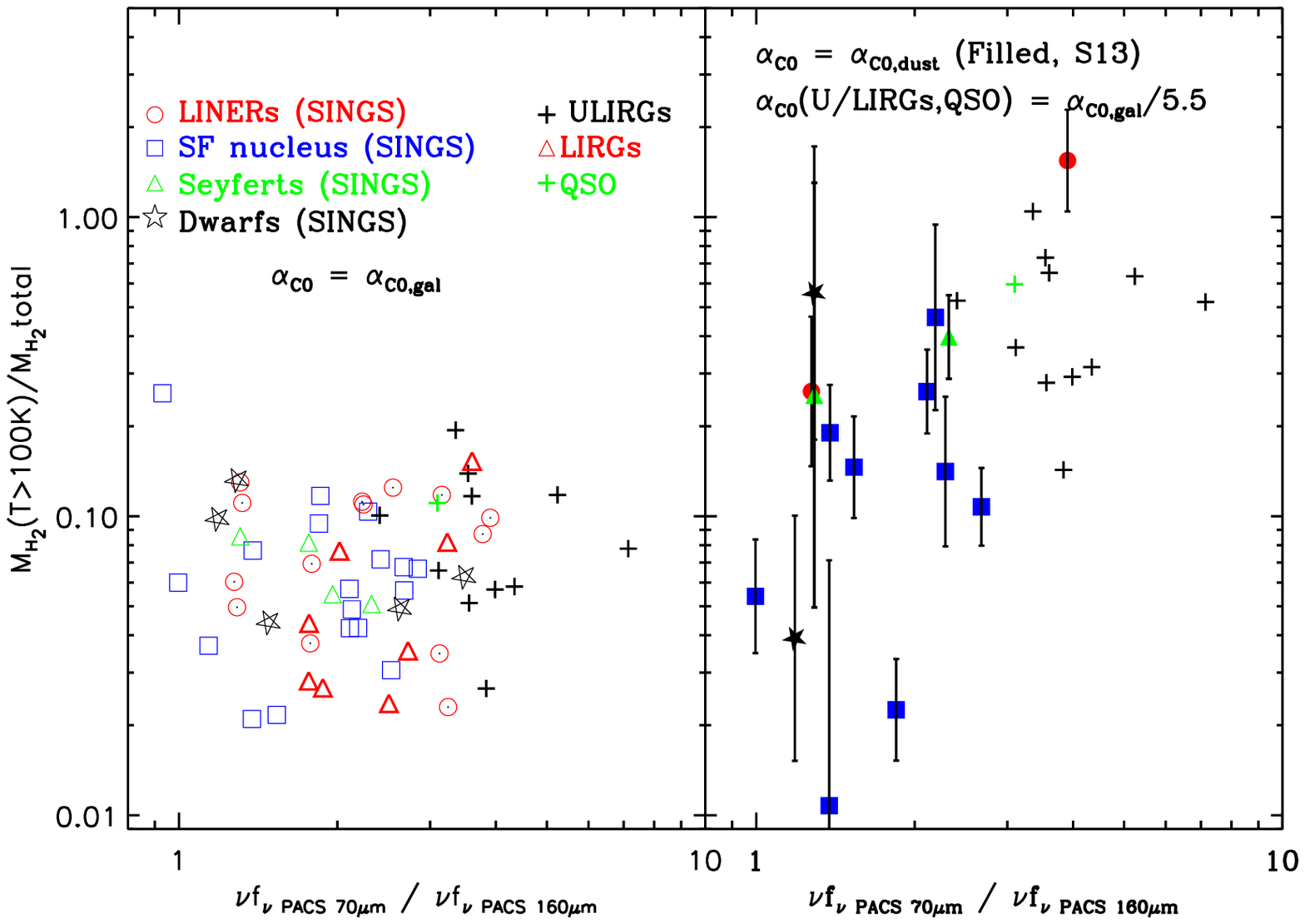}
\caption{The fraction of warm \htwo\ gas mass ($T>100$\,K) versus the
dust color temperature, $\rm{\nu f_{\nu 70 \micron}/\nu f_{\nu 160 \micron}}$, obtained from PACS.  At left a consistent Galactic $\rm{\alpha_{CO}}$ is adopted, and little trend is seen.  At right, a reduced $\rm{\alpha_{CO}}$ is applied to ULIRGs and QSO, and the dust-derived central $\rm{\alpha_{CO}}$ values of \citet{Sandstrom13} are used.  Galaxies with warmer dust color temperatures have high warm molecular gas mass fraction.}
\label{fig:wts}
\end{figure*}

Figure \ref{fig:wts} shows the fraction of warm molecular gas mass, $\rm{M(> 100 K)/M_{total}}$, as a function of far infrared (FIR) dust color temperature, $\rm{\nu f_{\nu 70 \micron}/\nu f_{\nu 160 \micron}}$. The warm gas fraction obtained using $\rm{\alpha_{CO,Gal}}$ ranges from 2--30\%, and exhibits little correlation among different galaxy types with dust color temperature. The ULIRGs and normal star forming galaxies show similar warm gas fractions, though ULIRGs have warmer dust color temperatures,  and LINERs and Seyferts have somewhat higher warm gas mass fractions than normal star forming galaxies on average.   

In contrast, using the available dust-derived central $\rm{\alpha_{CO}}$ estimates of \citet{Sandstrom13} for normal galaxies and a reduced value $\frac{1}{5.5}\times\rm{\alpha_{CO,Gal}}$ for ULIRGs and QSO's, however, leads to a strong correlation, with warmer dust implying an increasing warm \htwo\ fraction. The average warm gas fraction for ULIRGs and QSO is then $\sim$45\%, significantly above that of normal galaxies, and on the same increasing trend with dust temperature.  Depending on which prescription for total gas mass is correct, this could indicate a dependence of $\rm{\alpha_{CO}}$ on temperature.  

When considering a broken power law model, the fraction M($>$100 K)/M$_{total}$ is, as expected, very similar to that of a single power law for $T_{b} \leq$ 100 K. For  $T_{b} >$ 120\,K irrespective of low-temperature slope $n_{1}$, the model yields a poor fit to the excitation diagrams (see \S\,\ref{sec:broken-power-law}). At these intermediate break temperatures, the warm mass fraction  M($>$100 K)/M$_{total}$ changes by at most $\sim$10--20\% compared to the case of a single power law.

\subsection{Molecular gas in low metallicity galaxies}

As traced by their CO emission, many low metallicity dwarfs appear to have vanishingly low molecular gas content, but retain high star formation rates.  That is, they are strong outliers on the Schmidt-Kennicutt relation \citep{Galametz09, Schruba12}. This discrepancy implies either that in dwarf galaxies star formation efficiencies are higher compared to normal spirals, or they host large molecular gas reservoirs than is traced by the CO emission \citep{Schruba12, Glover12b}. It is possible that a significant fraction of \htwo\ exists outside the CO region, where the carbon is in C$^{+}$ (ionized) or C$^{0}$ (neutral) states. Since \htwo\ can self-shield from UV photons in regions where CO is photodissociated \citep{Wolfire10}, at low metallicity, not only is the CO abundance reduced, but as dust opacity is reduced and ionized regions become hotter and more porous, $\alpha_{\rm{CO, Gal}}$ can severely underestimate the molecular gas mass. Considerable effort has been invested in detecting and interpreting CO emission at metallicity 50 times lower than the solar metallicity, \twelveoh$\sim7.0$, to assess the molecular gas content \citep{Leroy11, Schruba12, Cormier14, Remy14}. Dust emission can be used to estimate the molecular gas mass in ISM, assuming a constant DGR however, it is essential to know the change in DGR with metallicity. At very low metallicities, \twelveoh$\le8$, DGR appears to scale non-linearly with metallicity \citep{Herrera12, Remy14}.

Our direct detection of \htwo\ gas mass through \htwo\ rotational lines is independent of any indirect tracers, which are affected by changing metallicity and local radiation effects. Applying our model in low metallicity galaxies should yield an estimate of molecular gas mass without the same inherent biases introduced by these dust and CO abundance variations. In this section we estimate the \htwo\ gas masses through our power law model in a low metallicity galaxy sample selected to have detected \htwo\ rotational emission, faint CO detection, and (where available) estimates of dust mass. We then compare these \htwo -based estimates to models and other methods which attempt to control for the biases introduced at low metallicity.

The low metallicity galaxies were selected on availability of MIR \htwo\ rotational lines to have atleast three rotational lines including S(0) or S(1) line fluxes along with CO derived molecular mass estimates. 

\subsubsection{Metallicity estimation}
To study the variation of \htwo\ gas mass from CO derived measurements over the metallicity range, it is essential to estimate the metallicity of galaxies. The metallicities were determined applying the direct $T_{e}$ method. CGCG\thinspace007-025 is the lowest metallicity galaxy in our dwarf sample with the value of \twelveoh = 7.77 \citep{Izotov07}. \citet{Guseva12} estimated the value of \twelveoh\ in the two H II regions, Haro\thinspace11B and Haro\thinspace11C as 8.1 and 8.33, respectively hence, we adopt the average value for Haro\thinspace11. The metallicity value, \twelveoh, for NGC\thinspace6822 is 8.2 \citep{Israel97}. No literature value exist for the oxygen gas phase abundance for the specific region of Hubble V in NGC\thinspace6822, mapped by the IRS-Spitzer. \citet{Peimbert05} estimated \twelveoh = 8.42$\pm$0.06 for Hubble V, which is inconsistent with the previous value of 8.2.

For selected SINGS galaxies, the metallicity values in the circumnuclear regions, which are approximately the size of our \htwo\ line flux extracted regions, are estimated by averaging the theoretical (KK04) and an empirical metallicity calibration (PTO5) as recommended by \citet{Moustakas10}.  

\subsubsection{Cold molecular gas from CO line emission}
Although the CO abundance drops super-lineraly with decreasing metallicity, it is detected in the low metallicity sample, and as the most common molecular tracer, can be compared directly to our \htwo\- based method. We adopted the literature values for $^{12}$CO(1--0) line intensities for CGCG\thinspace007-025 and N66, while for Haro\thinspace11, UM\thinspace311 and Hubble\thinspace V region $^{12}$CO(3--2) line intensities were scaled using the relation $\rm{I_{CO(3-2)}/I_{CO(1-0)}}$ = 0.60 (in temperature units) due to unavailability of $^{12}$CO(1--0) line intensities. Calculating $\rm{L\arcmin_{CO}}$, (area integrated luminosity) the molecular gas masses are estimated using $\rm{\alpha_{CO,Gal}}$ and were further scaled using the 8 $\micron$ map, to account for the difference in the extracted IRS spectrum and the CO beam regions for each galaxy. The molecular gas masses are listed in Table \ref{table:lowz}.

\subsubsection{Molecular gas from dust emission}
An alternative method for estimating molecular gas mass makes use of dust emission together with assumption of dust opacity and grain size distribution to calculate a total mass, scaling dust mass to the total gas mass using a presumed or modeled dust-to-gas ratio, and removing the measured atomic mass from the region.

\citet{Leroy07} estimated \htwo\ gas surface density using dust emission from FIR map in N66 region of SMC. They derived $\rm{\alpha_{CO}}$ to be about 27$\times\rm{\alpha_{CO,Gal}}$. For Haro\thinspace11, UM\thinspace311 and Hubble\thinspace V using metallicity-DGR relation from \citet{Sandstrom13} and with the known dust mass, we estimated the total gas mass and after subtracting the atomic gas content subsequently calculated the molecular gas mass. However, we find a negative value for the molecular gas for UM311, suggesting the ISM mass is mainly dominated by the atomic gas. The \htwo\ gas masses calculated from the dust emission are given in Table \ref{table:lowz}.

\subsubsection{Molecular gas mass using our model}
The \htwo\ line flux extraction was performed for the similar region in SMC-N66, where the CO emission was measured by \citet{Rubio96}. The cubes were prepared using CUBISM \citep{Smith07b} (Jameson, K. et al. in prep) and the \htwo\ line fluxes were estimated using the PAHFIT, a MIR spectral decomposition tool \citep{Smith07a}. The flux of \htwo\ rotational lines for Haro\thinspace11 are from \citet{Cormier14}, while for UM\thinspace311, and CGCG\thinspace007-025 \citet{Hunt10} measured the \htwo\ rotational line fluxes. For Hubble V region, \citet{Roussel07} derived the flux of \htwo\ rotational lines. 

\begin{figure}
\includegraphics*[width=0.5\textwidth]{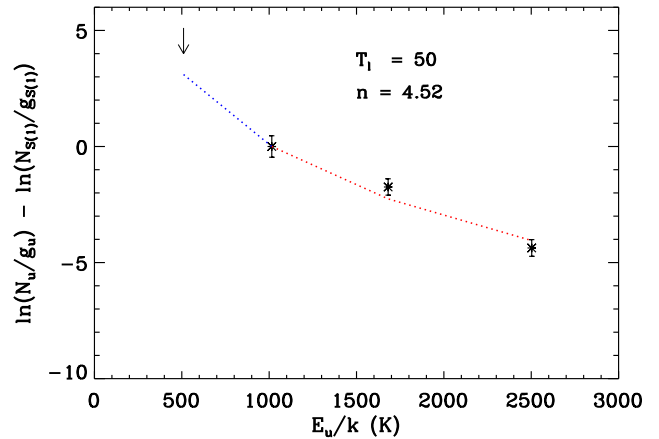}
\caption{Excitation diagram for low metallicity galaxy Haro\thinspace11. The N$_{u}$/g$_{u}$ ratios are normalized with respect to S(1) transition. The dashed red line indicate the model fit to the observed ratios. The blue dashed line predicts the value of S(0) flux ratio. The model estimated $T_{\ell}$, and n are mentioned.}
\label{fig:Haro}
\end{figure}

Figure \ref{fig:Haro} is an excitation diagram for low metallcity galaxy Haro\thinspace11, with the power law model fit. Our model prediction for the unobserved S(0) line is included, and is consistent with the estimated upper limit. The \htwo\ total gas mass for each low metallicity galaxy is measured using the power law model with lower temperature extrapolation to $T_{\ell}^{\star}$ = 49 K. Table \ref{table:lowz} lists the value of metallicity with their distance and the measured value of \htwo\ rotational line fluxes with the power law index and the \htwo\ gas mass derived using CO, dust, and our model.

\begin{figure*}
\centering
\includegraphics[width=0.95\textwidth]{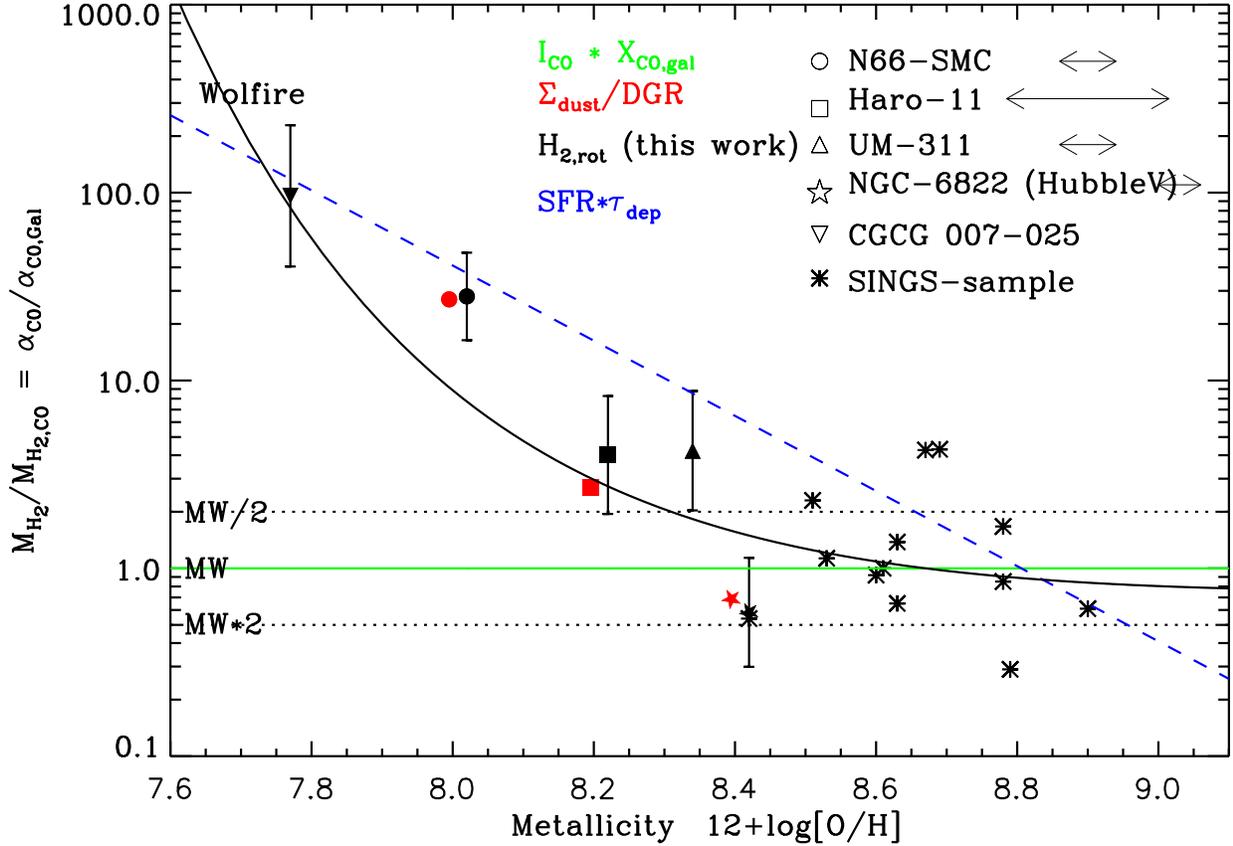}
\caption{The ratio of molecular gas masses estimated using different methods to the ``naive mass'' obtained using $\rm{L_{CO}}$ (and $\rm{\alpha_{CO}}$ = $\rm{\alpha_{CO,Gal}}$), as a function of gas-phase metallicity.  This ratio is equivalent to $\alpha_\mathrm{CO}/\rm{\alpha_{CO,Gal}}$. The black points show the molecular gas masses derived from the \htwo\ rotational model.  The blue line shows a fit to this ratio derived from inverting the star formation law for HERACLES non-starburst  galaxies \citep{Schruba12}. The molecular gas masses traced by the dust emission are denoted by the red points in the plot, and are shifted slightly in their metallicity values for clarity. The black solid line is the predicted mass ratio from the theoretical model of \citet{Wolfire10}, assuming $N({\textrm{H}_2})=10^{22}$ cm$^{-2}$ and $log(G_0/n)$ = -0.3. A steep increase in the ratio of model derived \htwo\ gas mass to the $\rm{L_{CO}}$ derived measurements is observed at metallicity values \twelveoh $\leq$ 8.4.}
\label{fig:lowZ}
\end{figure*}

Figure~\ref{fig:lowZ} compiles \htwo\ gas masses derived using the various indirect tracers, together with the results from our \htwo-only model.  All values are shown \emph{relative} to the \htwo\ mass inferred using CO luminosities with $\rm{\alpha_{CO} = \alpha_{CO,Gal}}$, and are plotted as a function of metallicity.  For the SINGS sample, a variation of about 2--3 times the $\rm{L_{CO}}$-derived \htwo\ gas mass is found, likely a consequence of the intrinsic variation in $\rm{\alpha_{CO}}$ at high metallicity  \twelveoh\ $\gtrsim 8.4$.  At intermediate metallicity, the molecular gas content from CO line emission for the Hubble V region in NGC\thinspace6822 compares well with our model-derived gas mass, which suggests a similar Galactic conversion factor ($\rm{\alpha_{CO}}$ = $\rm{\alpha_{CO,Gal}}$), consistent with the results of \citet{Rijcke06}. At lower metallicity, however, our model disagrees strongly with naive CO-based estimates, yielding up to $\sim100\times$ the molecular gas mass inferred from CO emission.

The molecular masses we recover at low metallicity are in good agreement with other measures which attempt to account for the impact of reduced metal abundance.  These include dust-derived measurements, where available (when the strong metallicity dependence of the dust-to-gass ratio is accounted for).  They also agree well with the prescription for the power-law like $\rm{\alpha_{CO}}$ variations with metallicity recovered from inverting star formation densities among all non-starburst galaxies in the HERACLES sample \citep{Schruba12}. The theoretical model of varying $\rm{\alpha_{CO}}$ by \citet{Wolfire10}, assuming \htwo\ column density of 10$^{22}$ cm$^{-2}$ in a molecular cloud, agrees well with observational results based on dust-mass at low metallicity \citep{Leroy11,Sandstrom13}.  Adopting a solar metallicity value of 8.66 and heating rate/H atom $\log(G_0/n)$ = -0.3 (from the $\rm{L_{OI}/L_{FIR}}$ studies of \cite{Malhotra01}), the \citet{Wolfire10} model also agrees well with our \htwo\ rotational emission modeled masses. 

The power law model reliably recovers the total molecular gas masses at metallicity as low as 10$\%$ of the Milky Way, where CO and other indirect tracers suffer strong and non linear biases.

\subsection{Future prospects}
At metallicities $\lesssim$ 0.25 Z$_{\odot}$, the direct power-law method recovers total molecular gas content as reliably as other tracers that account for or avoid the impact of reduced gas-phase metal content. At even lower metallicities, the CO abundance plummets, with essentially all of the molecular gas in a CO-dark phase. In the first epoch of the star formation in the universe, the extremely low abundance of heavy elements leaves \htwo\ as a principal coolant \citep{Lepp02}. 

The Mid Infrared Instrument (MIRI) onboard \emph{JWST} is sensitive enough to detect the S(1), S(2), S(3) and higher rotational lines of \htwo\ in luminous galaxies (U/LIRGs) till redshift 0.6, 1.3, 1.9, and higher, respectively. Assuming average
$\rm{L_{H_{2}S(1)}/L_{IR}}$ = 10$^{-4}$ \citep{Bonato15}, and $\rm{L_{IR}}$ = 3$\times$10$^{11}$ L$_{\odot}$ (typical for LIRGs), for S(1) at z = 0.5 to have signal to noise S/N = 5 will require 30 minutes of integration time with JWST-MIRI. It will be possible to measure pure \htwo\ rotational lines at high redshifts of z $\approx$ 6--7, almost reaching the reionization era of universe, with the SPace Infrared telescope for Cosmology and Astrophysics (SPICA) and the Cryogenic Aperture Large Infrared Submillimeter Telescope Observatory (CALISTO), planned for the 2020 decade  \citep{Roelfsema12, Bradford15}.

The above mentioned future projects for the next decade provides an opportunity to observe \htwo\ rotational lines at high redshifts. The power law model can be an useful tool in estimating molecular gas mass and study its variation and consequences at different redshifts.

\section{Summary}
We present a new power law temperature distribution model capable of reliably estimating the total molecular gas mass in galaxies purely from as few as three mid-infrared pure-rotational \htwo\ emission lines.  Our model is independent of the biases affecting indirect tracers like CO, dust emission, or star formation prescriptions.  It can hence be used even in environments where reliability of those indirect tracers is poor, such as at low metallicity.  We calibrate the model on a sample of local star-forming galaxies with well-quantified CO-based molecular gas masses, and apply the model to local ULIRGs and low metallicity systems.  Our key results are:

\begin{enumerate}
\item A model based on a continuous power law distribution of rotational temperatures reproduces well the \htwo\ excitation from a wide range of galaxy types using only a single parameter (the power law slope $n$).  This simple model can directly recover the warm \htwo\ mass ($T\!\gtrsim\!100$\,K) more reproducibly than arbitrary discrete temperature fits.

\item The power law index obtained for all SINGS galaxies ranges from 3.79 - 6.4, with an average value of 4.84 $\pm$ 0.61.

\item With typical Spitzer detection sensitivities, the model can directly recover the \htwo\ gas mass down to a limiting sensitivity temperature of $T_{s}=81$\,K (when the S(0) line is available), accounting for $\sim15\%$ of the total molecular reservoir.

\item By calibrating the model using a subset of the SINGS sample with well determined CO-based molecular masses, we find that extrapolating the \htwo\ temperature distribution to a \emph{single} cutoff temperature of $T_\ell^\star=49$\,K enables recovery of the total \htwo\ gas mass within a factor of 2.2 (0.34 dex).

\item Evaluating a family of broken power law models, we find that the average lower cutoff temperature $T_{\ell,b}^{\star}$ could range from 5--50 K, while recovering total \htwo\ gas masses as well as a single power law model. However, low model slopes below the break ($n_1$)  would require unphysically low cutoff temperatures, and for all power law indices $n_1\leq3$, break temperatures $T_{b}\gtrsim$120\,K result in unacceptable fits to the \htwo\ MIR rotational lines.

\item When $\rm{\alpha_{CO,Gal}}$ is used, the fraction of warm molecular gas mass (M$>$ 100 K) in this training sample ranges from 0.02 to 0.30.  If a reduced $\rm{\alpha_{CO}}$ is adopted for the ULIRGs, this fraction increases with increasing dust color temperature 
$\rm{\nu f_{\nu 70 \micron}/\nu f_{\nu 160 \micron}}$.

\item In ULIRGs, the total molecular gas mass obtained by extrapolating the model to $T_\ell^\star=49$\,K is consistent with the molecular gas mass derived using $\rm{\alpha_{CO,Gal}}$. Alternatively, if a reduced $\rm{\alpha_{CO}}$ is adopted, the model extrapolation temperature required rises to 80 K.  Either the warm molecular gas fraction and lower temperature cutoff in ULIRGs is higher than in normal star forming galaxies, or $\rm{\alpha_{CO}}$ is closer to the Galactic value than has been presumed.

\item At low metallicity ($12+\log[\textrm{O}/\textrm{H}]\!\lesssim\!8.4$), where indirect tracers of molecular mass suffer increasingly strong and non-linear biases and the mass of the molecular reservoirs exceed their CO-derived estimates by $100\times$ or more, the direct power law model recovers the total molecular gas masses reliably, as assessed by other methods which account for metallicity.  

\item With upcoming facilities including \emph{JWST}, \emph{SPICA}, and \emph{CALISTO}, detection of the relevant \htwo\ rotational lines in galaxies at intermediate to high redshifts becomes possible, opening a new window on the fueling history of star formation in the Universe. 
\end{enumerate}

Acknowledgements: This paper is dedicated to James R. Houck of Cornell University (1940-2015), the PI of the {\it Spitzer IRS}. This work is based on observations made with the \emph{Spitzer Space Telescope}. We have made use of the NASA/IPAC Extragalactic Database (NED) which is operated by the Jet Propulsion Laboratory, California Institute of Technology under NASA contract. Support for this research was provided by NASA through contract 1287374 issued by JPL/Caltech under contract 1407. We thank the anonymous referee for useful comments and suggestions. We are grateful to K. Jameson and A. Bolatto for sharing the N66-SMC region spectrum.  We are also indebted to L. Armus, N. Scoville, B. Draine, A. Bolatto, S. R. Federman, K. Sandstrom, D. Neufeld, F. Walter, T. Diaz-Santos, E. Pellegrini for discussions which improved this work, and to S. Glover for providing model access and for helpful discussions of simulated \htwo\ temperature in galaxies.  JDS gratefully acknowledges visiting support from the Alexander von Humboldt Foundation and the Max Planck Institute f\"{u}r Astronomie, and support from a Cottrell Scholar Award from the Research Corporation for Science Advancement.  AT acknowledges support for this work through the IPAC Visiting Graduate Fellowship program.”

\bibliography{plainnat}

\begin{longtable}{cccccc}
\tabletypesize{\scriptsize}
\tablecaption{Observed properties of our sample galaxies}
\tablewidth{0pt}
\tablehead{
\colhead{Galaxy} & \colhead{D}  & \colhead{Type} & \colhead{L$_{IR}$}   & \colhead{$\frac{S_{70}}{S_{160}}$} & \colhead{Ref}\\
\colhead{Name}     & \colhead{(Mpc)} & \colhead{} & \colhead{(10$^{10}$ L$_{\odot}$)}& \colhead{} & \colhead{}\\
(1) & (2) & (3) & (4) & (5) & (6)
}
N0337 & 19.3 & SF  & 0.568 & 0.807 & S07\\
N1097 & 14.2 & SF  & 2.343 & 0.932 & S07\\
N1266 & 31.0 & LIN & 1.333 & 1.420 & S07\\
N1291 & 10.4 & LIN & 0.039 & 1.382 & S07\\
N1316 & 21.0 & LIN & 0.260 & 1.368 & S07\\
N1482 & 22.6 & SF  & 2.812 & 1.169 & S07\\
N1566 & 20.4 & SY  & 0.444 & 0.772 & S07\\
N2798 & 25.8 & SF  & 2.316 & 0.407 & S07\\
N2976 & 3.6  & DW  & 0.007 & 0.564 & S07\\
N3049 & 19.2 & SF  & 0.307 & 1.107 & S07\\
N3184 & 11.4 & SF  & 0.026 & 0.604 & S07\\
N3190 & 19.3 & LIN & 0.290 & 0.577 & S07\\
N3198 & 14.1 & SF  & 0.147 & 0.813 & S07\\
N3265 & 19.6 & SF  & 0.237 & 1.057 & S07\\
Mrk33 & 22.9 & DW  &\nodata &1.525 & --\\
N3351 & 9.3  & SF  & 0.268 & 0.924 & S07\\
N3521 & 11.2 & LIN & 0.266 & 0.572 & S07\\
N3627 & 6.4  & SY  & 0.221 & 1.016 & S07\\
N3938 & 14.3 & SF  & 0.105 & 0.436 & S07\\
N4125 & 23.9 & LIN & 0.048 & 0.975 & S07\\
N4254 & 14.4 & SF  & 0.533 & 0.602 & S07\\
N4321 & 14.3 & SF  & 0.513 & 0.671 & S07\\
N4450 & 16.5 & LIN & 0.092 & 1.116 & S07\\
N4536 & 14.5 & SF  & 0.905 & 1.172 & S07\\
N4559 & 7.00 & SF  & 0.083 & 0.498 & S07\\
N4569 & 16.8 & LIN & 0.426 & 0.777 & S07\\
N4579 & 16.4 & SY  & 0.161 & 0.857 & S07\\
N4625 & 7.6  & DW  & 0.023 & 0.518 & S07\\
N4631 & 9.3  & SF  & 0.306 & 0.922 & S07\\
N4725 & 20.5 & SY  & 0.034 & 0.564 & S07\\
N4736 & 4.7  & LIN & 0.115 & 1.708 & S07\\
N4826 & 5.3  & LIN & 0.116 & 0.782 & S07\\
N5033 & 14.8 & SY  & 0.577 & 0.572 & S07\\
N5055 & 7.9  & LIN & 0.159 & 0.557 & S07\\
N5194 & 7.6  & SY  & 0.308 & 0.686 & S07\\
N5195 & 7.6  & LIN & 0.154 & 1.652 & S07\\
N5713 & 21.4 & SF  & 2.998 & 1.001 & S07\\
N5866 & 15.3 & LIN & 0.215 & 0.981 & S07\\
N6822A& 0.5  & DW  & \nodata &1.151& --\\
N6946 & 6.8  & SF  & 0.387 & 0.958 & S07\\
N7331 & 14.5 & LIN & 0.249 & 0.564 & S07\\
N7552 & 21.0 & SF  & 5.052 & 1.244 & S07\\
N7793 & 3.9  & DW  & 0.009 & 0.648 & S07\\
3c31    & 66.7 & 3C & \nodata & \nodata & \nodata\\
3c218   & 240  & 3C & \nodata & \nodata & \nodata\\
3c272.1 & 19.1 & 3C & \nodata & \nodata & \nodata\\
3c293   & 195  & 3C & \nodata & \nodata & \nodata\\
3c310   & 233  & 3C & \nodata & \nodata & \nodata\\
3c326n  & 395  & 3C & \nodata & \nodata & \nodata\\
3c424   & 568  & 3C & \nodata & \nodata & \nodata\\
3c433   & 445  & 3C & \nodata & \nodata & \nodata\\
3c436   &1016  & 3C & \nodata & \nodata & \nodata\\
CenA    & 11.0 & 3C & \nodata & \nodata & \nodata\\
3c236   & 449  & 3C & \nodata & \nodata & \nodata\\
Arp220          & 77.6& ULI & 145 & 1.678   & H06\\
IRAS 00188-0856 & 596 & ULI & 256 & \nodata & H06\\
IRAS 03521+0028 & 717 & ULI & 365 & \nodata & H06\\
IRAS 05189-2524 & 186 & ULI & 143 & 2.292   & H06\\
IRAS 06035-7102 & 356 & ULI & 166 & \nodata & H06\\
IRAS 06206-6315 & 418 & ULI & 169 & \nodata & H06\\
IRAS 07598+6508 & 700 & ULI & 337 & \nodata & H06\\
IRAS 08572+3915 & 258 & ULI & 137 & 3.124   & H06\\
IRAS 10565+2448 & 188 & ULI & 109 & 1.362   & H06\\
F12112+0305     & 324 & ULI & 212 & 1.576   & H06\\
IRAS 13451+1232 & 563 & ULI & 197 & \nodata & H06\\
F14348-1447     & 332 & ULI & 224 & 1.550   & H06\\
IRAS 17208-0014 & 188 & ULI & 250 & 1.744   & H06\\
IRAS 19254-7245 & 273 & ULI & 121 & 1.468   & H06\\
IRAS 20087-0308 & 483 & ULI & 280 & \nodata & H06\\
IRAS 23365+3604 & 286 & ULI & 147 & 1.557   & H06\\
Mrk 273         & 164 & ULI & 142 & 1.899   & H06\\
UGC5101         & 174 & ULI & 100 & 1.053   & H06\\
Mrk 463E        & 221 & ULI & 60 &  \nodata & H06\\
PG1440+356      & 347 & QSO & 42 & 1.355    & E01\\
N6240           & 101 & LIR & 69 & 1.415    & H06\\
N3110           & 72.5& LIR & 16 & 0.772    & P10\\
N3256           & 40.1& LIR & 40 & 1.577    & P10\\
N3690           & 44.9& LIR & 63 & 1.663    & P10\\
N5135           & 58.8& LIR & 16 & 1.096    & P10\\
N6701           & 56.2& LIR & 10 & 0.883    & P10\\
N7130           & 69.9& LIR & 25 & 1.192    & P10\\
N7591           & 70.9& LIR & 10 & 0.821    & P10\\
N7771           & 62.1& LIR & 25 & 0.773    & P10\\

\tablenotetext{}{1. Ref: S07 - Smith et al. 2007; K09 - Kennicutt et al. 2009; NED - NASA/IPAC Extragalactic Database; H06 - Higdon et al. 2006; P10 - Pereira Santaella et al. 2010; E01 - Evans et al. 2001}
\tablenotetext{}{2. Entries in the last column are references for the the infrared luminosity (8-1000 $\micron$), $\rm{L_{IR}}$. For SINGS galaxies $\rm{L_{IR}}$ is calculated using the relation $\rm{L_{IR} = 0.94 \times L_{TIR}}$ (D. Dale priv. comm.) and the corresponding $\rm{L_{TIR}}$ values are from \cite{Smith07a}.}
\tablenotetext{}{3. Distance measurements are from the KINGFISH webpage for SINGS galaxies and for others through NED IPAC Extragalactic Database}
\tablenotetext{}{4. The $\rm{S_{70}/S_{160}}$ is the FIR ratio calculated from the PACS 70 and 160 $\micron$ fluxes}
\tablenotetext{}{5. SF: star forming region; LIN: LINERs; SY: Seyferts; DW: Dwarfs; 3C: 3C radio galaxies; ULI: ULIRGs; QSO: Quasar; and LIR: LIRGs}
\tablenotetext{}{6. The 3c radio galaxies were selected from \cite{Ogle10}.}
\label{table:sample}
\end{longtable}
\newpage
\begin{longtable*}{cccccccccc}
\tabletypesize{\scriptsize}
\tablecaption{Observed molecular hydrogen rotational line flux}
\tablewidth{0pt}
\tablehead{
\colhead{Galaxy} & \colhead{S(0)(err)} & \colhead{S(1)(err)} & \colhead{S(2)(err)} & \colhead{S(3)(err)} & \colhead{S(4)(err)} & \colhead{S(5)(err)} & \colhead{S(6)(err)} & \colhead{S(7)(err)} & \colhead{M(H$_{2}$)} \\
\colhead{Name} &
\multicolumn{8}{c}{---------------------------------------------------------10$^{-17}$ W m$^{-2}$---------------------------------------------------------} &
\colhead{(10$^{6}$M$\odot$)}\\
(1) & (2) & (3) & (4) & (5) & (6) & (7) & (8) & (9) & (10)
}
N0337\tablenotemark{1,1} & 01.16(0.38) & 02.10(0.49) & 01.54(0.99) & 00.97(0.52) & \nodata & \nodata & \nodata & \nodata & 56 (228) \\
N1097\tablenotemark{1,1} & 21.31(4.26) & 72.61(4.33) & 29.36(2.94) & 42.30(2.34) & \nodata & \nodata & \nodata & \nodata & 1493 \\
N1266\tablenotemark{25,1} & 01.64(0.85) & 14.85(0.66) & 12.18(0.71) & 18.98(1.16) & 10.27(1.74) & 24.19(1.99) & \nodata & 19.22(2.86) & 1253 \\
N1291\tablenotemark{1,1} & 00.37(0.18) & 02.98(0.48) & 01.39(0.82) & 03.16(1.00) & \nodata & \nodata & \nodata & \nodata & 9 \\
N1316\tablenotemark{1,1} & 00.15(0.08) & 03.60(0.61) & 01.84(0.59) & 08.40(0.99) & \nodata & \nodata & \nodata & \nodata & 124 \\
N1482\tablenotemark{1,1} & 10.68(3.51) & 42.40(5.34) & 18.40(2.28) & 20.75(2.69) & \nodata & \nodata & \nodata & \nodata & 1607 \\
N1566\tablenotemark{26,1} & 02.40(0.22) & 12.95(0.71) & 05.53(0.48) & 09.16(1.76) & \nodata & \nodata & \nodata & \nodata & 389\\
N2798\tablenotemark{1,1} & 04.32(2.53) & 20.78(2.21) & 09.07(0.92) & 10.52(1.21) & \nodata & \nodata & \nodata & \nodata & 273 \\
N2976\tablenotemark{1S,1} & 00.89(0.24) & 01.57(0.30) & 00.49(0.35) & 00.52(0.42) & \nodata & \nodata & \nodata & \nodata & 1.08 (0.25) \\
N3049\tablenotemark{1,1} & 00.64(0.22) & 2.37(0.38) & 01.17(0.65) & 01.11(0.52)  & \nodata & \nodata & \nodata & \nodata & 148 \\
N3184\tablenotemark{1S,1} & 00.98(0.23) & 02.26(0.29) & 00.63(0.35) & \nodata & \nodata & \nodata & \nodata & \nodata & 26 (10.6)\\
N3190\tablenotemark{1,1} & 01.81(0.29) & 07.53(0.80) & 02.09(0.64) & 07.16(1.26) & \nodata & \nodata & \nodata & \nodata & 110 \\
N3198\tablenotemark{1,1} & 01.32(0.49) & 03.21(0.54) & 01.02(0.35) & 01.87(0.85) & \nodata & \nodata & \nodata & \nodata & 31 \\
N3265\tablenotemark{1,1} & 00.93(0.32) & 03.06(0.35) & 01.14(0.79) & 01.53(0.74) & \nodata & \nodata & \nodata & \nodata & 94 \\
Mrk33\tablenotemark{1,1} & 01.39(0.56) & 02.93(0.37) & 01.15(0.53) & 04.57(0.86) & \nodata & \nodata & \nodata & \nodata & 81 \\
N3351\tablenotemark{1S,1} & 06.28(1.05) & 21.84(2.06) & 08.83(1.01) & 16.98(1.99) & \nodata & \nodata & \nodata & \nodata & 197 (32) \\
N3521\tablenotemark{1,1} & 01.78(0.36) & 03.12(0.69) & 01.23(0.40) & 02.05(0.88) & \nodata & \nodata & \nodata & \nodata & 20  \\
N3627\tablenotemark{1S,1} & 03.12(0.39) & 31.86(1.22) & 14.17(0.52) & 20.91(1.26) & \nodata & \nodata & \nodata & \nodata & 208 (27) \\
N3938\tablenotemark{1S,1} & 00.80(0.12) & 01.34(0.33) & 00.42(0.32) & \nodata & \nodata & \nodata & \nodata & \nodata & 37 (30)\\
N4125\tablenotemark{1,1} & 00.26(0.16) & 01.76(0.47) & 01.26(0.64) & 01.80(0.88) & \nodata & \nodata & \nodata & \nodata & 26 \\
N4254\tablenotemark{1S,1} & 01.76(0.13) & 08.73(0.95) & 04.66(0.92) & 03.25(0.88) & \nodata & \nodata & \nodata & \nodata & 342 (700) \\
N4321\tablenotemark{1S,1} & 07.82(1.17) & 26.75(2.28) & 13.21(2.46) & 16.45(1.62) & \nodata & \nodata & \nodata & \nodata & 1386 (200)\\
N4450\tablenotemark{1,1} & 01.03(0.16) & 09.14(0.52) & 03.37(0.95) & 08.90(1.11) & \nodata & \nodata & \nodata & \nodata & 63 \\
N4536\tablenotemark{1S,1} & 10.87(3.23) & 41.16(4.05) & 17.56(2.33) & 21.68(2.07) & \nodata & \nodata & \nodata & \nodata & 810 (415) \\
N4559\tablenotemark{1,1} & 01.36(0.21) & 01.79(0.35) & 00.40(0.15) & 02.34(1.37) & \nodata & \nodata & \nodata & \nodata & 18 \\
N4569\tablenotemark{1,1} & 04.16(0.69) & 31.61(0.87) & 15.24(0.52) & 30.94(1.87) & \nodata & \nodata & \nodata & \nodata & 760  \\
N4579\tablenotemark{1,1} & 00.89(0.23) & 16.41(0.52) & 10.58(1.24) & 25.63(1.00) & \nodata & \nodata & \nodata & \nodata & 205 \\
N4625\tablenotemark{1S,1} & 00.80(0.16) & 00.96(0.19) & 00.54(0.43) & \nodata & \nodata & \nodata & \nodata & \nodata & 8 (21)\\
N4631\tablenotemark{1,1} & 05.82(0.64) & 12.36(1.61) & 06.04(0.50) & 04.68(0.91) & \nodata & \nodata & \nodata & \nodata & 73.4  \\
N4725\tablenotemark{1S,1} & 01.17(0.17) & 3.71(0.51) & 1.99(0.81) & 3.80(0.90) & \nodata & \nodata & \nodata & \nodata & $<$170 ($<$28) \\
N4736\tablenotemark{1S,1} & 03.32(0.68) & 25.08(1.64) & 10.01(1.04) & 22.17(1.36) & \nodata & \nodata & \nodata & \nodata & 21 (1.4) \\
N4826\tablenotemark{1,1} & 07.13(0.65) & 34.53(1.58) & 15.09(1.09) & 21.06(1.03) & \nodata & \nodata & \nodata & \nodata & 65 \\
N5033\tablenotemark{1,1} & 03.66(0.35) & 18.20(1.04) & 06.35(0.31) & 12.69(1.91) & \nodata & \nodata & \nodata & \nodata & 190 \\
N5055\tablenotemark{1S,1} & 04.40(0.24) & 15.80(0.91) & 05.20(0.67) & 08.02(1.06) & \nodata & \nodata & \nodata & \nodata & 87 (20)\\
N5194\tablenotemark{1,1} & 01.77(0.28) & 13.44(0.91) & 07.57(0.47) & 18.73(1.77) & \nodata & \nodata & \nodata & \nodata & 32 \\
N5195\tablenotemark{1,1} & 05.42(2.09) & 31.04(1.23) & 12.77(0.65) & 27.50(1.72) & \nodata & \nodata & \nodata & \nodata & 80 \\
N5713\tablenotemark{1S,1} & 02.74(0.36) & 15.16(1.43) & 05.09(0.79) & 08.56(0.70) & \nodata & \nodata & \nodata & \nodata & 329 (238) \\
N5866\tablenotemark{1,1} & 01.55(0.15) & 09.00(0.55) & 03.91(0.70) & 03.72(0.65) & \nodata & \nodata & \nodata & \nodata & 81 \\
N6822A\tablenotemark{1,1} & 00.81(0.43) & 02.11(0.43) & 01.60(0.41) & 02.59(0.38) & \nodata & \nodata & \nodata & \nodata & 0.03 \\
N6946\tablenotemark{1S,1} & 22.77(6.19) & 63.69(3.35) & 27.16(1.71) & 28.63(2.93) & \nodata & \nodata & \nodata & \nodata & 346 (32) \\
N7331\tablenotemark{1,1} & 01.64(0.16) & 04.96(0.50) & 01.38(0.19) & 02.34(0.83) & \nodata & \nodata & \nodata & \nodata & 120  \\
N7552\tablenotemark{1,1} & 25.78(10.69)& 56.53(4.26) & 24.59(2.25) & 31.59(1.48) & \nodata & \nodata & \nodata & \nodata & 1950 \\
N7793\tablenotemark{1,1} & 00.70(0.17) & 01.50(0.32) & 01.17(0.46) & 00.53(0.37) & \nodata & \nodata & \nodata & \nodata & 3 \\
3c031\tablenotemark{2,8} & 00.64(0.11)  & 01.36(0.58) & 00.59(0.15) & 0.60(0.17)  & 00.76(0.27) & 00.80(0.18) & \nodata & \nodata  & 695 \\
3c218\tablenotemark{2,9} & 00.55(0.11)  & 00.41(0.22) & 00.23(0.09) & 00.55(0.08) & 00.42(0.15) & 00.23(0.05) & \nodata & 0.16(0.05) & 1895 \\
3c272.1\tablenotemark{2,10} & 00.42(0.14) & 00.55(0.29) & 00.27(0.15) & 01.40(0.20) & \nodata & 01.00(0.40) & 01.60(0.40) & \nodata & 1.8 \\
3c293\tablenotemark{2,11} & 01.60(0.20)  & 05.30(0.40) & 01.94(0.07) & 03.25(0.07) & 01.20(0.10) & 02.80(0.20) & 00.31(0.11) & 1.30(0.10) & 14450 \\
3c310\tablenotemark{2} & 00.12(0.06) & 00.36(0.07) & 00.08(0.04) & 00.48(0.04) & 00.19(0.06) & \nodata & \nodata & 00.29(0.09) & \nodata \\
3c326n\tablenotemark{2,12} & 00.30(0.06) & 00.69(0.06) & 00.41(0.04) & 01.26(0.05) & 00.31(0.06) & 00.46(0.22) & 00.25(0.09) & 00.29(0.09) & 1183 \\
3c424\tablenotemark{2,13} & 00.45(0.07) & 01.12(0.06) & 00.21(0.03) & 00.43(0.04) & 00.15(0.04) & 00.17(0.07) & \nodata & 00.28(0.05) & $<$3560 \\
3c433\tablenotemark{2,14} & 02.00(0.40) & 01.40(0.20) & 00.58(0.17) & 01.20(0.20) & \nodata & 01.60(0.50) & \nodata & 01.20(0.30) & $<$5495 \\
3c436\tablenotemark{2} & 00.48(0.16) & 00.46(0.07) & 00.31(0.12) & 00.28(0.04) & \nodata & 00.22(0.10) & 00.41(0.12) & 00.21(0.10) & \nodata \\
CenA\tablenotemark{2,15} & 10.00(6.00) & 57.00(9.00) & 27.00(4.00) & 09.30(3.70) & \nodata & 12.00(3.00) & \nodata & 27.00(4.00) & $<$210 \\
3c236\tablenotemark{3,10} & 00.20(0.03) & 00.95(0.07) & 00.32(0.05) & 00.64(0.05) & \nodata & \nodata & \nodata & \nodata & $<$6220 \\
Arp220\tablenotemark{4,16} & $<970$ & 18.62(1.68) & 09.80(1.30) & 07.30(0.20) & \nodata & \nodata & \nodata & 12.90(3.80) & 22500 \\
IRAS 00188-0856\tablenotemark{4,16} & $<19$ & 00.43(0.05) & 00.20(0.04) & 00.38(0.12) & \nodata & \nodata & \nodata & $<32$ & 21300\\
IRAS 03521+0028\tablenotemark{4,16} & $<13$ & 00.82(0.13) & 00.27(0.09) & 00.37(0.03) & \nodata & \nodata & \nodata & $<3.0$ & 29400\\
IRAS 05189-2524\tablenotemark{4,16} & $<282$ & 03.40(0.70) & 01.50(0.20) & 03.60(1.30) & \nodata & \nodata & \nodata & $<341$ & 3200\\
IRAS 06035-7102\tablenotemark{4,17} & $<61$ & 04.17(0.04) & 02.33(0.58) & 03.4(1.00) & \nodata & \nodata & \nodata & $<116$ & 19000\\
IRAS 06206-6315\tablenotemark{4,17} & $<29$ & 01.29(0.10) & 00.50(0.05) & 00.59(0.19) & \nodata & \nodata & \nodata & $<18$ & 41400 \\
IRAS 07598+6508\tablenotemark{4,16} & $<30$ & 01.38(0.70) & 00.37(0.06) & 00.67(0.02) & \nodata & \nodata & \nodata & $<225$ & 36000 \\
IRAS 08572+3915\tablenotemark{4,16} & $<154$ & 01.19(0.04) & 00.51(0.10) & 00.46(0.13) & \nodata & \nodata & \nodata & $<430$ & 4200 \\
IRAS 10565+2448\tablenotemark{4,16} & $<109$ & 06.57(0.05) & 02.56(0.09) & 04.03(0.02) & \nodata & \nodata & \nodata & $<54$ & 17900\\
F12112+0305\tablenotemark{4,18} & 1.4(0.4) & 4.12(0.44) & 1.73(0.31) & 2.37(0.03) & \nodata & \nodata & \nodata & 4.10(0.8) & 19200 \\
IRAS 13451+1232\tablenotemark{4,18} & $<45$ & 02.83(0.65) & 01.29(0.05) & 02.19(0.26) & \nodata & \nodata & \nodata & $<41$ & $<$46900 \\
F14348-1447\tablenotemark{4,18} & $<6.1$ & 4.47(0.15) & 1.95(0.10) & 2.4(0.21) & \nodata & \nodata & \nodata & $<31$ & 20500 \\
IRAS 17208-0014\tablenotemark{4,16} & $<239$ & 08.81(0.09) & 04.97(0.85) & 05.7(1.1) & \nodata & \nodata & \nodata & $<85$ & 22400 \\
IRAS 19254-7245\tablenotemark{4,17} & $<85$ & 08.81(0.57) & 03.82(0.95) & 03.82(0.04) & \nodata & \nodata & \nodata & $<4.00$ & 21000\\
IRAS 20087-0308\tablenotemark{4,16} & $<21$ & 02.29(0.14) & 00.84(0.33) & 01.3(0.40) & \nodata & \nodata & \nodata & $<22$ & 51100 \\
IRAS 23365+3604\tablenotemark{4,16} & $<91$ & 03.9(0.31) & 02.01(0.95) & 03.1(0.5) & \nodata & \nodata & \nodata & $<30$ & 27000 \\
Mrk 273\tablenotemark{4,16} & $<262.5$ & 10.24(0.09) & 05.6(0.70) & 10.4(0.9) & \nodata & \nodata & \nodata & $<150$ & 15870 \\
UGC5101\tablenotemark{4,19} & $<100$ & 04.96(0.47) & 02.7(0.5) & 02.8(0.30) & \nodata & \nodata & & $<195$ & 7500\\
Mrk463E\tablenotemark{4,18} & $<79$ & 02.79(0.37) & 01.31(0.08) & 02.8(0.60) & \nodata & \nodata & & $<385$ & 2180\\
PG1440+356\tablenotemark{5,20} & 0.60(0.14) & 1.14(0.14) & 0.51(0.06) & 0.64(0.11) & \nodata & \nodata & \nodata & \nodata & 6400 \\
N6240\tablenotemark{6,21} & 8.60(1.20) & 50.40(1.50) & 39.8(0.50) & 70.90(1.90) & 36.4(10.8) & 95.2(16.6) & \nodata & 33.7(12.5) & 14100 \\
N3110\tablenotemark{7,21} & 1.60(0.20) & 10.00(0.90) & 4.00(1.00) & \nodata & \nodata & \nodata & \nodata & \nodata & 5750 \\
N3256\tablenotemark{7,22} & 9.00(3.00) & 61.0(5.00) & 34.0(3.00) & \nodata & \nodata & \nodata & \nodata & \nodata & 1400 \\
N3690\tablenotemark{7,21} & 3.80(0.60) & 22.0(1.00) & 11.0(4.00) & \nodata & \nodata & \nodata & \nodata & \nodata & 3700 \\
N5135\tablenotemark{7,21} & 1.30(0.50) & 17.00(1.00) & 9.30(1.40) & \nodata & \nodata & \nodata & \nodata & \nodata & 5500 \\
N6701\tablenotemark{7,21} & 1.90(0.90) & 10.00(1.00) & 4.60(0.80) & \nodata & \nodata & \nodata & \nodata & \nodata & 1200 \\
N7130\tablenotemark{7,23} & 1.40(0.30) & 09.90(0.70) & 4.20(1.70) & \nodata & \nodata & \nodata & \nodata & \nodata & 4130 \\
N7591\tablenotemark{7,24} & 1.00(0.30) & 05.50(0.60) & 3.00(1.00) & \nodata & \nodata & \nodata & \nodata & \nodata & 3270 \\
N7771\tablenotemark{7,21} & 3.50(1.80) & 16.00(1.00) & 8.70(1.20) & \nodata & \nodata & \nodata & \nodata & \nodata & 3280 \\
\tablenotetext{}{1. The cold H$_{2}$ gas mass is calculated using $\rm{\alpha_{CO,Gal}}$ = 3.2 M$_\odot$(K km s$^{-1}$ pc$^{2}$)$^{-1}$ (not considering Helium and other heavy element mass contribution)} 
\tablenotetext{}{2. The mass value in the parentheses is calculated using CO weighted mean $\rm{\alpha_{CO}}$, suggested by Sandstrom et al. (Table 4) 2013 using a constant dust to gas ratio (DGR)  marked in the reference as ``S''}
\tablenotetext{}{3.The two numbers as the notemark in the first column of the table correspond to references. The first number is the reference from which we adopt the MIR \htwo\ rotational line flux and the second number is for the cold \htwo\ gas mass derived from the CO emission.\\
Ref: (1) Roussel et al. 2007; (1S) Sandstrom et al.2013 + Roussel et al. 2007; (2) Ogle et al. 2010; (3) Guillard et al. 2012; (4) Hiqdon et al. 2006; (5) QUEST sample by S. Veilleux (2009); (6) Armus et al. 2006; (7) M. Pereira-Santaella. 2010; (8) Okuda et al. 2005; (9) Salome $\&$ Combes 2003; (10) Oca{\~n}a Flaquer et al. 2010; (11) Evans et al. 1999; (12) Nesvadba et al. 2010; (13) Saripalli $\&$ Mack 2007; (14) Evans et al. 2005; (15) Israel et al. 1990; (16) Solomon et al. 1997; (17) Mirabel et al. 1989; (18) Evans et al. 2002; (19) Rigopoulou 1996; (20) Evans et al. 2001; (21) Sanders et al. 1991; (22) Kazushi Sakamoto et al. (2006); (23) Curran et al. (2000); (24) Lavezzi $\&$ Dickey 1998; (25) Young et al. (2011); (26) Harnett et al.(1991)}
\label{table:h2flux}
\end{longtable*}
\newpage
\begin{deluxetable}{ccccc}
\centering
\tablewidth{0pt}
\tablecolumns{5}
\tablecaption{Observed molecular hydrogen lines}
\tablehead{
\colhead{Transition} &
\colhead{Short notation} &
\colhead{Rest $\lambda$} &
\colhead{$\frac{E_{u}}{k}$} &
\colhead{A} \\
\colhead{$\nu=0$} &
\colhead{} &
\colhead{($\micron$)} &
\colhead{(K)} &
\colhead{($10^{-11} s^{-1}$)}\\
(1) & (2) & (3) & (4) & (5)
}
\startdata
$J=2\longrightarrow0$ & S(0) & 28.219 & 0510 & 2.95\\
$J=3\longrightarrow1$ & S(1) & 17.035 & 1015 & 47.6\\
$J=4\longrightarrow2$ & S(2) & 12.279 & 1681 & 275.0\\
$J=5\longrightarrow3$ & S(3) &  9.665 & 2503 & 980.0\\
$J=6\longrightarrow4$ & S(4) &  8.025 & 3473 & 2640.0\\
$J=7\longrightarrow5$ & S(5) &  6.910 & 4585 & 5880.0\\
$J=8\longrightarrow6$ & S(6) &  6.109 & 5828 & 11400.0\\
$J=9\longrightarrow7$ & S(7) &  5.511 & 7196 & 20000.0\\
\enddata
\tablecomments{The rotational upper level energies were computed from the molecular constants given by \cite{Huber79} and transition probabilities are from \cite{Black76}
}
\label{table:lineppt}
\end{deluxetable}

\begin{deluxetable}{cccc}
\centering
\tablecaption{Model derived parameters for SINGS galaxies}
\tablewidth{0pc}
\tablecolumns{4}
\tablehead{
\colhead{Galaxy}  & 
\colhead{$T_{\ell}$} & \colhead{n}   & \colhead{$\frac{M(T>100 K)}{M_{total}}$}\\
\colhead{Name}    & 
\colhead{(K)} & \colhead{} & \colhead{(in $\%$)}\\
(1) & (2) & (3) & (4)
}
\startdata
N0337      & 59 (43)     & 5.47 & 9.46 (2.3)\\
N1097      & 47          & 5.00 & 4.88       \\
N1266      & 26          & 3.80 & 2.30       \\ 
N1291      & 52          & 4.27 & 11.79      \\
N1316      & 30          & 3.79 & 3.48       \\
N1482      & 51          & 5.00 & 6.76       \\
N1566      & 42          & 3.89 & 8.15       \\
N2798      & 71          & 4.96 & 25.76      \\
N2976      & 66 (89)     & 5.87 & 13.22 (56.6)\\
N3049      & 42          & 5.02 & 3.06       \\
N3184      & 57 (69)     & 5.57 & 7.67 (18.3)\\
N3190      & 55          & 4.68 & 11.08       \\
N3198      & 60          & 5.20 & 11.70   \\
N3265      & 53          & 5.15 & 7.17     \\
Mrk33      & 43          & 4.28 & 6.28      \\
N3351      & 43 (70)     & 4.75 & 4.22 (26.2)\\
N3521      & 63          & 5.42 & 12.97      \\
N3627      & 43 (77)    & 4.53 & 5.08 (39.7) \\
N3938      & 57 (55)     & 6.00 & 6.02 (5.03)\\
N4125      & 50          & 4.16 &11.18\\
N4254      & 35 (29)     & 4.68 & 2.10 (1.05) \\
N4321      & 38 (62)     & 4.96 & 2.18 (15.1)\\
N4450      & 53          & 4.28 & 12.46\\
N4536      & 49 (58)     & 5.03 & 5.64 (11.1)\\
N4559      & 51          & 5.90 & 3.69\\
N4569      & 37          & 4.30 & 3.76 \\
N4579      & 37          & 3.92 & 5.48\\
N4625      & 65 (54)     & 6.39 & 9.81 (3.61)\\
N4631      & 51          & 5.25 & 5.72 \\
N4725      & $>51$       & 4.79 & $<$7.8 \\
N4736      & 52 (112)    & 4.54 & 9.88 (149)\\
N4826      & 50          & 4.85 & 6.93\\
N5033      & 51          & 4.65 & 8.56\\
N5055      & 50 (72)     & 5.05 & 6.04 (26.4)\\
N5194      & 39          & 3.94 & 6.28\\
N5195      & 50          & 4.51 & 8.78\\
N5713      & 56 (61)     & 4.91 & 10.36 (14.5)\\
N5866      & 55          & 4.70 & 10.95\\
N6822A     & 39          & 4.20 & 4.91\\
N6946      & 45 (82)     & 4.96 & 4.23 (45.6)\\
N7331      & 49          & 5.21 & 4.96 \\
N7552      & 51          & 5.02 & 6.67\\
N7793      & 48          & 5.25 & 4.42
\enddata
\tablecomments{The value in the parentheses is calculated assuming central $\alpha_{CO}$ from the dust emission, evaluated by Sandstrom et al. 2013}
\label{table:modelvalue}
\end{deluxetable}
\begin{deluxetable*}{cccccccc}
\centering
\tablecaption{Model derived parameters for radio, U/LIRGs galaxies}
\tablewidth{0pc}
\tablecolumns{9} 
\tablehead{
\colhead{Galaxy}  & \colhead{Ref} & \colhead{Model mass} & \colhead{$T_{\ell}$} & \colhead{n}   & \colhead{$\frac{M(T>100 K)}{M_{total}}$} & \colhead{$T'_{\ell}$} & \colhead{$\frac{M(T>100 K)}{M_{total}}$}\\
\colhead{Name}    & \colhead{}    & \colhead{($10^{6}M_{\odot}$)} & \colhead{(K)} & \colhead{} & \colhead{(in $\%$, for $T_{\ell}$)} & \colhead{(K)} & \colhead{(in $\%$ for $T'_{\ell}$)}\\
(1) & (2) & (3) & (4) & (5) & (6) & (7) & (8)
}
\startdata
3c031           & 1  & 360      & 41       & 4.69  & 3.73   & 65     & 20.40\\
3c218           & 1  & 1040     & 41       & 4.47  & 4.53   & 67     & 24.92\\
3c272.1         & 1  & 1.87     & 50       & 3.41  & 18.8   & 101    & 102.4\\
3c293           & 1  & 13210    & 48       & 4.77  & 6.28   & 75     & 33.81\\
3c310           & 1  & 515      & \nodata  & 4.11  &\nodata &\nodata &\nodata\\
3c326n          & 1  & 2604     & 64       & 4.05  & 25.0   & 112    & 126.2\\
3c424           & 1  & 40820    & $>$88    & 5.19  &$>$58.5 & $>$132 & $>$320.0\\
3c433           & 1  & 12850    & $>$62    & 4.51  &$>$18.7 & $>$101 & $>$103.6\\
3c436           & 1  & 18820    & \nodata  & 4.40  &\nodata &\nodata &\nodata \\
CenA            & 1  & 379      & $>$58    & 4.64  &$>$13.8 & $>$93  & $>$76.79\\
3c236           & 2  & 14083    & $>$61    & 4.86  &$>$14.8 & $>$95  & $>$82.04\\
Arp220          & 3  & 10000    & 41     & 5.07  & 2.65  & 62 & 14.13\\
IRAS 00188-0856 & 3  & 5472     & 33     & 4.38  & 2.36  & 55 & 13.13\\
IRAS 03521+0028 & 3  & 53970    & 57     & 5.35  & 8.67  & 84 & 48.02\\
IRAS 05189-2524 & 3  & 3600     & 52     & 4.27  & 11.8  & 87 & 62.94\\
IRAS 06035-7102 & 3  & 19240    & 50     & 4.39  & 9.54  & 82 & 51.55\\
IRAS 06206-6315 & 3  & 11760    & 36     & 4.68  & 2.33  & 56 & 11.80\\
IRAS 07598+6508 & 3  & 76700    & 60     & 5.27  & 11.3  & 88 & 58.93\\
IRAS 08572+3915 & 3  & 6586     & 53     & 5.02  & 7.80  & 85 & 51.64\\
IRAS 10565+2448 & 3  & 19746    & 51     & 5.04  & 6.60  & 78 & 35.82\\
IRAS 12112+0305 & 3  & 38640    & 59     & 5.07  & 11.7  & 90 & 63.80\\
IRAS 13451+1232 & 3  & 43890    &$>$49   & 4.57  & $>$7.83 &$>$78 &$>$41.27\\
IRAS 14348-1447 & 3  & 37130    & 59     & 4.74  & 13.9  & 92 & 72.38\\
IRAS 17208-0014 & 3  & 14900    & 45     & 4.59  & 5.69  & 71 & 29.40\\
IRAS 19254-7245 & 3  & 89760    & 69     & 5.42  & 19.4 & 101 & 106.48\\
IRAS 20087-0308 & 3  & 45390    & 49     & 5.03  & 5.64  & 74 & 28.89\\
IRAS 23365+3604 & 3  & 18230    & 45     & 4.72  & 5.13  & 71 & 27.30\\
Mrk 273         & 3  & 9730     & 43     & 4.37  & 5.82  & 71 & 31.55\\
UGC5101         & 3  & 11655    & 56     & 4.96  & 10.1  & 85 & 53.17\\
Mrk 463E        & 3  & 4320     & 61     & 4.24  & 19.7  & 102 & 107.1\\
PG1440+356      & 4  & 11735    & 58     & 5.04  & 11.1  & 88 & 59.58\\
N6240           & 5  & 7240     & 40     & 3.73  & 8.20  & 65 & 30.95\\
N3110           & 5  & 1520     & 33     & 4.22  & 2.82  & 51 & 11.33\\
N3256           & 5  & 1360     & 50     & 3.71  & 15.3  & \nodata & \nodata\\
NGC3690/IC694           & 5  & 1310     & 36     & 4.04  & 4.48  & 56 & 17.21\\
N5135           & 5  & 820      & 25     & 3.70  & 25.2  & 41 & 9.17\\
N6701           & 5  & 800      & 44     & 4.13  & 7.65  & 68 & 30.36\\
N7130           & 5  & 1320     & 35     & 4.18  & 3.55  & 54 & 14.13\\
N7591           & 5  & 754      & 32     & 4.18  & 2.67  & 49 & 10.17\\
N7771           & 5  & 1000     & 33     & 3.82  & 4.39  & 54 & 17.25
\enddata
\tablecomments{1. The molecular gas mass in column 3 is calculated extrapolating the power law model to T$_{\ell}^{\star} = 49$ K.\\
2. The temperature in column 4 is the model extrapolated temperature required to fit the total molecular gas mass, calculated using the $\rm{\alpha_{CO,Gal}}$.\\
3. The warm gas mass fraction listed in column 6 is derived using the cold molecular gas mass by assuming $\rm{\alpha_{CO,Gal}}$.\\
4. The temperature in column 7 is the model extrapolated temperature required to fit the total molecular gas mass assuming a low $\rm{\alpha_{CO}}$ of 0.8 M$_\odot$(K km s$^{-1}$ pc$^{2}$)$^{-1}$.\\
5. The warm gas mass fraction listed in column 8 is derived using the cold molecular gas mass by assuming a low $\rm{\alpha_{CO}}$}
\label{table:uli}
\end{deluxetable*}
\begin{table}
\rotatebox{90}{
\begin{tabular}{ccccccccccccc}
\centering
 & \multicolumn{11}{c}{\textbf{Table 6}} & \\
 & \multicolumn{11}{c}{{Observed molecular hydrogen line fluxes and mass in low metallicity dwarfs}} & \\
\\
\hline
\hline
Galaxy & 12+log[O/H]$^{a}$ & S(0) & S(1) & S(2) & S(3) & S(4) & S(5) & D & M($\rm{H_{2}}$,CO)$^{c}$ & M($\rm{H_{2}}$,dust)$^{d}$ & M($\rm{H_{2}}$,model) & n\\
Name &  & \multicolumn{6}{c}{--------------------------------------10$^{-17}$ W m$^{-2}$--------------------------------------} & Mpc & 
\multicolumn{3}{c}{-----------------10$^{5}$ M$_{\odot}$-----------------} & \\
(1) & (2) & (3) & (4) & (5) & (6) & (7) & (8) & (9) & (10) & (11) & (12) & (13)\\
\hline
CG007-025$^{b}$ & 7.77 & \nodata & 5.45(0.43) & 1.67(0.43) & \nodata & 0.72(0.21) & \nodata & 24.5 & 41.2 & \nodata & 3935 & 5.24\\
N66 & 8.10 & 475(181)&1025(116) & 1283(175)& 433(207) & 586(400)& 482(314)& 0.06 & 0.016 & 0.435 & 0.451 & 3.59\\
Haro 11 & 8.20 & $<$2.49 & 1.68(0.55) & 1.01(0.13) & 1.21(0.18) & \nodata & \nodata & 92 & 1670 & 4500$^{e}$ & 6700 & 4.52\\
UM311   & 8.31 & \nodata & 1.49(0.69) & \nodata &  \nodata  & 0.73(0.35) & 0.66(0.27) & 24 & 95 & \nodata & 400 & 4.51\\
HubbleV & 8.42 & 0.81(0.43) & 2.11(0.43) & 1.60(0.41) & 2.59(0.38) & \nodata & \nodata & 0.5 & 0.27 & 0.28 & 0.16 & 4.20\\
\hline
\end{tabular}
}
\rotatebox{90}{
\centering
a. Ref for metallicity CGCG\thinspace007-025:\cite{Izotov07}; N66:\cite{Dufour82,Dufour84}; Haro\thinspace11:\cite{Guseva12}; UM\thinspace311:\cite{Izotov98};}
\rotatebox{90}{
\centering
and HubbleV:\cite{Peimbert05}}
\rotatebox{90}{
\centering
b. For CGCG\thinspace007-025 we also used the S(7) line flux of H$_{2}$, $\rm{(0.46\pm0.13) \times 10^{-17} W m^{-2}}$, from \cite{Hunt10}}
\rotatebox{90}{
\centering
c. Ref for CO derived \htwo\ mass CGCG\thinspace007-025:\cite{Hunt15}; N66:\cite{Rubio96}; Haro\thinspace11 and UM\thinspace311:\cite{Cormier14}; HubbleV:\cite{Israel03}}
\rotatebox{90}{
\centering
d. Ref for dust derived \htwo\ mass N66:\cite{Leroy07}; Haro11:\cite{Cormier14}; HubbleV:\cite{Israel96}}
\rotatebox{90}{
\centering
e. The molecular gas mass for Haro11 is in the range (4.5--11.5)$\times10^{8}$ M$_{\odot}$, due to uncertainty in HI mass. In the table we mention the minimum value}
\label{table:lowz}
\end{table}
\end{document}